\documentclass[%
 reprint,
superscriptaddress,
 amsmath,amssymb,
pra,
]{revtex4-2}

\usepackage{graphicx}% Include figure files
\usepackage{dcolumn}% Align table columns on decimal point
\usepackage{bm}% bold math
\usepackage{amsmath}
\begin{document}

\preprint{APS/123-QED}

\title{Unified theory for frequency combs in ring and Fabry-Perot quantum cascade lasers: an order-parameter equation approach}%

\author{Carlo Silvestri}
\affiliation{School of Electrical Engineering and Computer Science, The University of Queensland, Brisbane, QLD 4072, Australia
}
\author{Massimo Brambilla}
\affiliation{ Dipartimento Interateneo di Fisica, Politecnico di Bari and CNR-IFN, UOS Bari, Italy
}
\author{Paolo Bardella}%
\affiliation{Dipartimento di Elettronica e Telecomunicazioni, Politecnico di Torino, 10129 Torino, Italy
}
\author{Lorenzo Luigi Columbo}
\affiliation{Dipartimento di Elettronica e Telecomunicazioni, Politecnico di Torino, 10129 Torino, Italy
}

\begin{abstract}
We present a unified model to describe the dynamics of optical frequency combs (OFCs) in quantum cascade lasers (QCLs), incorporating both ring and Fabry-Pérot (FP) cavity configurations. The model derives  a modified complex Ginzburg-Landau equation (CGLE), leveraging an order parameter approach and is capable of capturing the dynamics of both configurations, thus enabling a comparative analysis. In the modified CGLE, a nonlinear integral term appears which is associated with the coupling between counterpropagating fields in the FP cavity and whose suppression yields the ring model, which is known to be properly described by a conventional CGLE. We show that this crucial term holds a key role in inhibiting the formation of harmonic frequency combs (HFCs), associated to multi-peaked localized structures, due to its anti-patterning effect. We provide in support a comprehensive campaign of numerical simulations, in which we observe a higher occurrence of HFCs in the ring configuration compared to the FP case. Furthermore, the  simulations demonstrate the model's capability to reproduce experimental observations, including the coexistence of amplitude and frequency modulation, linear chirp, and typical dynamic scenarios observed in QCLs. Finally, we perform a linear stability analysis of the single-mode solution for the ring case, confirming its consistency with numerical simulations and highlighting its predictive power regarding the formation of harmonic combs.
\end{abstract}
\maketitle

%\tableofcontents

\section{Introduction}
Quantum cascade lasers (QCLs) stand as a pivotal technology in the realm of semiconductor lasers \cite{Faist_1994,Kohler2002}, distinguished, among other features, by their capability to spontaneously emit optical frequency combs (OFCs) across the mid-infrared (mid-IR) and terahertz (THz) spectral ranges \cite{Faist_2016,Hugi2012,Burghoff2014,PiccardoReview}. Since their initial demonstration in 2012 \cite{Hugi2012}, the study of these OFCs has progressed rapidly, drawing a keen interest from both theoretical and experimental perspectives \cite{SilvestriReview,PiccardoReview}. 

The spontaneous formation of OFCs in QCLs has been thoroughly examined for the ring and the Fabry-Perot (FP) configurations. Remarkably, experimental techniques such as Shifted Wave Interference Fourier Transform Spectroscopy (SWIFTS) and Fourier transform analysis of comb emission (FACE) have enabled the retrieval of periodic temporal profiles of amplitude and instantaneous frequency of the electric field, shedding light on the coexistence of amplitude modulated (AM) and frequency modulated (FM) behavior within these combs \cite{Burghoff2014,BurghoffSWIFTS,Optica_Faist_18,Bomeng1,Cappelli2019,Faist_comb_profile_2023}. Another significant achievement is the experimental demonstration of spontaneous harmonic frequency combs (HFCs), where the comb spacing is a multiple of the laser cavity's free-spectral range (FSR), in both configurations \cite{PiccardoHFCOptex, Dhillon1,ForrerHFC,Kazakov2017,Kazakov2021}.

%This progress includes the demonstration and characterization of several types of these regimes and the unraveling of their origin and the physical mechanisms giving rise to them \cite{PiccardoReview,SilvestriReview}. The spontaneous formation of combs in QCLs has been extensively studied for both cavity configurations, namely the ring and the FP. From an experimental standpoint, a remarkable result is the retrieval of the periodic temporal profile of both amplitude and instantaneous frequency of the electric field for these OFCs, by exploiting Shifted Wave Interference Fourier Transform Spectroscopy (SWIFTS) \cite{Burghoff2014,BurghoffSWIFTS,Optica_Faist_18,Bomeng1} or Fourier transform analysis of comb emission (FACE) \cite{Cappelli2019} techniques. This has revealed the coexistence of amplitude modulated (AM) and frequency modulated (FM) behaviour for these combs \cite{Optica_Faist_18, Cappelli2019,Burghoff2014, Faist_comb_profile_2023}. Another significant accomplishment is the experimental demonstration of spontaneous harmonic frequency combs (HFCs), where the comb spacing is a multiple of the laser cavity's free-spectral range (FSR), in both configurations \cite{PiccardoHFCOptex, Dhillon1,ForrerHFC,Kazakov2017,Kazakov2021}.\\
%Furthermore, significant effort has been devoted to the theoretical investigation of comb formation in QCLs and the identification of the key physical mechanisms responsible for it, progressing alongside the experimental findings \cite{SilvestriReview}. 
Parallel efforts have been directed toward theoretical investigations to elucidate the underlying physical mechanisms that govern the formation of combs in QCLs \cite{SilvestriReview}. Spatial hole burning (SHB) and linewidth enhancement factor (LEF, also named $\alpha$ factor) have emerged as key factors that trigger multimode dynamics close to the laser threshold \cite{PiccardoSHB,PiccardoReview,Silvestri20,Opacakalpha, Boiko1,Boiko2}. Although both elements are present and collaborate to generate multimode emission in the FP configuration, in the unidirectional ring cavity SHB does not occur and therefore the LEF serves as the physical mechanism responsible for multimode dynamics \cite{NaturePiccardo, Columbo2018}. Then, the strong nonlinearity inherent in QCL heterostructures plays a crucial role in the proliferation and locking of generated optical lines \cite{friedli,Opacak2019}, competing with and compensating for the group velocity dispersion (GVD) arising from material and waveguide dispersion \cite{Opacak2019}.

Early theoretical studies utilized Maxwell-Bloch equations (MBEs) to reproduce and characterize QCL comb generation, although lacking inclusion of the LEF \cite{Khurgin2014,Tzenov2016, Jiraus1, Tzenov2, Boiko1,Boiko2}, and preventing then the theoretical reproduction of the near-threshold single-mode instability observed experimentally in ring configurations \cite{NaturePiccardo, Bomeng1}. To address this limitation, effective semiconductor Maxwell-Bloch equations (ESMBEs) were introduced for ring \cite{Columbo2018} and FP \cite{Silvestri20} QCLs. Derived self-consistently from a semiconductor optical susceptibility, ESMBEs incorporate key semiconductor properties, including the non-zero LEF. This model accurately reproduces typical characteristics in the time and frequency domains for both THz \cite{Silvestri22,Columbo2018} and mid-IR \cite{Silvestri20, Opacak2019} QCL combs.
%The initial theoretical studies on QCL comb generation utilized Maxwell-Bloch equations (MBEs), yielding specific comb characteristics, including frequency modulation and typical spectra replication \cite{Khurgin2014,Tzenov2016,Boiko1,Boiko2}. However, MBEs lack the inclusion of the LEF, preventing the theoretical reproduction of near-threshold single-mode instability observed experimentally in ring configurations \cite{NaturePiccardo, Bomeng1}. Multimode behavior can only be captured well above the threshold with ring MBEs due to the Risken-Nummedal-Graham-Haken instability \cite{Risken}. To address this, effective semiconductor Maxwell-Bloch equations (ESMBEs) were introduced for ring \cite{Columbo2018} and FP \cite{Silvestri20} QCLs.Derived self-consistently from a semiconductor optical susceptibility, ESMBEs incorporate key semiconductor properties, including the non-zero LEF. This model accurately reproduces typical characteristics in time and frequency domains for both THz \cite{Silvestri22,Columbo2018} and mid-IR \cite{Silvestri20} QCL combs, including simultaneous AM and FM behavior, linear chirp, variation between combs and unlocked regimes with varying pump parameters, and the formation of HFCs.\

However, the mathematical complexity of the MBEs and ESMBEs hinders analytical treatment and the identification of dominant mechanisms responsible for specific regimes, such as linear chirp or HFC emission. Reduced models, which are based on fewer equations and possess a lower mathematical complexity, have been shown to be more suitable for this purpose \cite{Opacak2019, Burghoff20, Humbard22, SilvestriThesis, NaturePiccardo, NB2024}.

In this work, we present a unified theoretical framework for describing the dynamics of QCL combs, unifying the dynamics of ring and FP QCLs into a single spatiotemporal equation using an order parameter approach proposed for Kerr combs and solitons in\cite{Cole2018,Oppo2023}. The equation takes the form of a modified complex Ginzburg-Landau equation (CGLE) for an auxiliary field, incorporating a non-local integral term accounting for the coupling between counterpropagating fields in the FP configuration due to SHB. We highlight that this modified CGLE is equivalent to two coupled CGLEs for the forward and backward fields propagating inside the FP cavity. Through suppression of the integral term, thus neglecting the field coupling, we can configure the model for the unidirectional ring case, described by a conventional CGLE for the unidirectional field, aligning seamlessly with previous investigations on this configuration \cite{NaturePiccardo, Prati2021,Pratichaos, Columbo2021,NB2024}.

This approach not only offers a unified and concise description of the dynamics in both systems, but also enables simultaneous characterization of the amplitude and phase dynamics of the field. This capability results in the possibility to reproduce the coexistence between AM and FM features observed in experimental setups. This represents an advance in the theoretical depiction of these combs compared to previous theories based on a single spatiotemporal equation \cite{Burghoff20}, which focused primarily on reproducing the typical FM behavior of QCL OFCs while neglecting the amplitude dynamics. Furthermore, the proposed model allows for a comparative analysis between the two configurations, highlighting differences in the formation of localized structures associated with the presence or absence of the non-local integral term.

Another advantage offered by the model is the possibility to conduct systematic simulations, involving scans across the pump parameter, which enable the reproduction of dynamic scenarios in accordance with experiments. This includes capturing key characteristics of QCL combs, such as the linear chirp and typical temporal profiles, as well as the generation of harmonic states. Furthermore, we emphasize that the reduction of the number of equations governing the dynamics of the two configurations holds significant promise for deducing general properties, gaining insight into physical phenomena, and conducting analytical treatments, such as deriving the linear stability analysis (LSA) of the single-mode solution.

In Section~\ref{Sec_deriv}, we outline the derivation of the two coupled CGLEs for the FP case from a full set of ESMBEs, and then we retrieve the single CGLE for the ring configuration as a special case.

In Section~\ref{secderivsingleeq}, we derive the single spatiotemporal equation for the order parameter, while in Section~\ref{secroleK}, we analyze the practical implications of the nonlinear integral term in the formation of multiple localized structures per round trip, corresponding to harmonic combs.

Section \ref{sec_num} is dedicated to the numerical results obtained by integrating the reduced models for both the ring (single CGLE) and the FP cases (two coupled CGLEs), enabling a comparison between the two configurations and the replication of several experimentally demonstrated features of the QCL combs.

In Section \ref{sec_LSA} we develop the LSA of the single-mode solution for the ring case, and we verify the consistency between the prediction of the LSA and the numerical simulations.

Section \ref{sec_concl} draws the conclusion of the work.
\section{Derivation of the reduced models for FP and ring QCL}\label{Sec_deriv}
\subsection{Two coupled Complex Ginzburg-Landau equations for the FP configuration}
We start from a full set of ESMBEs for the FP configuration, obtained by introducing scaling of the variables into the original equations presented in \cite{Silvestri20}:
%\begin{widetext}
\begin{eqnarray}
   \frac{\partial F^+}{\partial \eta}+  \frac{\partial F^+}{\partial t'} = \sigma\left[ -F^+-p^+\right] \label{el_+5}\\
    -  \frac{\partial F^-}{\partial \eta}+  \frac{\partial F^-}{\partial t'} = \sigma\left[ -F^--p^-\right] \label{el-_5}
    \end{eqnarray}
    \begin{equation}
        \frac{\partial p^+}{\partial t'}=\Gamma(1+i\alpha)[-p^+ -\left(1+i\alpha\right)\left(D_0F^{+}+D_1^+F^-\right)] \label{p+_5}
    \end{equation}
        \begin{equation}
        \frac{\partial p^-}{\partial t'}=\Gamma(1+i\alpha)[-p^- -\left(1+i\alpha\right)\left(D_0F^{-}+D_1^-F^+\right)] \label{p-_5}
    \end{equation}
    \begin{eqnarray}
     \frac{\partial D_0}{\partial t'}&=&  b\left[\mu-D_0+F^{+*}p^++F^{-*}p^-+F^{+}p^{+*}+F^{-}p^{-*}\right]\nonumber\\ \label{d0_5}  \\
        \frac{\partial D_1^+}{\partial t'}&=&  b\left[-D_1^++F^{-*}p^-+F^{+}p^{-*}\right] \label{d1_5}
\end{eqnarray}
where $\eta$ and $t^{\prime}$ represent the dimensionless scaled space and time variables, $F^+$ and $F^-$ denote the forward and backward envelopes of the electric fields, $p^+$ and $p^-$ are the forward and backward polarization terms, $D_0$ stands for the zero-order density of carriers, and $D_1^+$ and $D_1^-$ are the variables associated with the carrier grating due to SHB. Additionally, $\alpha$ represents the LEF, $\Gamma$ is an adimensional constant proportional to the gain linewidth (see \cite{Columbo2018}), $\sigma$ is the ratio between the polarization dephasing time $\tau_\mathrm{d}$ and the photon lifetime $\tau_\mathrm{p}$, while $b$ stands for the ratio between $\tau_\mathrm{d}$ and the carrier lifetime $\tau_\mathrm{e}$.

Eqs.~(\ref{el_+5})-(\ref{d1_5}) are completed with the boundary conditions for the FP cavity:
\begin{eqnarray}
F^-(L^\prime,t^\prime)&=&\sqrt{R}F^+(L^\prime,t^\prime),\label{bcc1}\\
F^+(0,t^\prime)&=&\sqrt{R}F^-(0,t^\prime),\label{bcc2}
\end{eqnarray}
where $L^\prime$ is the scaled cavity length, and $R$ is the reflectivity of the QCL facets.\\
One can observe that the term $p^+$ in the field equation Eq.~(\ref{el_+5}) is preceded by a negative sign, and similarly for $p^-$ in Eq.~(\ref{el-_5}). We remark that this is solely attributed to the chosen scaling. In fact, in the original unscaled ESMBEs \cite{Silvestri20}, the polarization terms act as sources in the field equations, as expected. Further details are provided in the Supplementary Material, where the original ESMBEs and the introduced scaling procedure are presented in detail.

We introduce the smallness parameter:
\begin{equation}
    \epsilon=\sqrt{\sigma}
\end{equation}
and we assume fast carriers and near threshold operation. Therefore we can write:
\begin{eqnarray}
   F^\pm&=&\varepsilon\ F^{(1)\pm}+O(\varepsilon^2) \label{camposvil}\\ 
p^\pm&=&\varepsilon\ p^{(1)\pm}+O(\varepsilon^2)\\
D_0&=&1+\varepsilon^2D_0^{(2)}+O(\varepsilon^3)\\ 
{D_1}^\pm&=&\varepsilon^2\ {D_1}^{(2)\pm}+O(\varepsilon^3)\label{reticolosvil}\\
\mu&=&1+\varepsilon^2\mu^{(2)}+O(\varepsilon^3) \label{musvil}
\end{eqnarray}
We have introduced the notation $X^{(n)}$ to denote the $n$-th order term in the expansion of the variable $X$. The expansion of $\mu$ in Eq.~(\ref{musvil}) corresponds to the implementation of the hypothesis of a near-threshold operation. Furthermore, in order to have derivatives of order O(1), we assume that the following Taylor expansions hold:
\begin{eqnarray}
   \frac{\partial}{\partial t^\prime}&=&\ \frac{\partial}{\partial{t^\prime}^{(0)}}+\varepsilon^2\frac{\partial}{\partial{t^\prime}^{(2)}}+O(\varepsilon^3)\label{expans1}\\
\frac{\partial}{\partial\eta}&=& \frac{\partial}{\partial\eta^{(0)}}+\varepsilon^2\frac{\partial}{\partial\eta^{(2)}}+O(\varepsilon^3)\label{expans2}
\end{eqnarray}
By introducing the expansion Eq.~(\ref{expans1}), we can rewrite the equation Eq.~(\ref{el_+5}) for the forward field:
\begin{multline}
\left.\ \varepsilon\frac{\partial F^{\left(1\right)+}}{\partial\eta^{(0)}}+\varepsilon^3\frac{\partial F^{\left(1\right)+}}{\partial\eta^{(2)}}\right.+\left.\ \varepsilon\frac{\partial F^{\left(1\right)+}}{\partial{t^\prime}^{(0)}}+\varepsilon^3\frac{\partial F^{\left(1\right)+}}{\partial{t^\prime}^{(2)}}\right.=\\\varepsilon^2\left[-\varepsilon\ F^{\left(1\right)+}-\varepsilon\ p^{\left(1\right)+}\right]  \label{field01}
\end{multline}
At first order in $\epsilon$ we have:
\begin{equation}
\left.\ \frac{\partial F^{\left(1\right)+}}{\partial\eta^{(0)}}=\right.\left.\ -\frac{\partial F^{\left(1\right)+}}{\partial{t^\prime}^{(0)}}\right.\label{uguag}
\end{equation}
Now let us consider the equation for $p^+$, Eq.~(\ref{p+_5}) and let us introduce the expansions Eqs.~(\ref{expans1})-(\ref{expans2}) into it. We have at first order in $\epsilon$:
\begin{equation}
    \varepsilon\frac{\partial p^{\left(1\right)+}}{\partial{t^\prime}^{(0)}}=\mathrm{\Gamma}(1+i\alpha)\left\{-\varepsilon\ p^{\left(1\right)+}-\left(1+i\alpha\right)\varepsilon\ F^{\left(1\right)+}\right\}\
\end{equation}
and then:
\begin{equation}
    \left[1+\frac{1}{\mathrm{\Gamma}(1+i\alpha)}\frac{\partial}{\partial{t^\prime}^{(0)}}\right]\varepsilon p^{\left(1\right)+}=-\left(1+i\alpha\right)\varepsilon\ F^{\left(1\right)+}\ \label{P1ord}
\end{equation}
If we solve Eq.~(\ref{P1ord}) in the Fourier domain we obtain:
\begin{equation}
    \left[1+\frac{i\omega}{\mathrm{\Gamma}(1+i\alpha)}\right]{\hat{p}}^{\left(1\right)+}=-\left(1+i\alpha\right)\ {\hat{F}}^{\left(1\right)+}
\end{equation}
By introducing the additional hypothesis that $\omega/\Gamma<<1$: 
\begin{equation}
    p^{\left(1\right)+}=-\left(1+i\alpha\right)\ F^{\left(1\right)+}\label{ppp}
\end{equation}
Using Eq.~(\ref{ppp})  we get from Eqs.~(\ref{d0_5})-(\ref{d1_5}):
\begin{eqnarray}
   {D_1}^{(2)\pm}&=&-2F^{\left(1\right)\mp\ast}F^{\left(1\right)\pm}\\
   D_0&=&\mu-2\left(|F^+|^2+|F^-|^2\right)
\end{eqnarray}
We solve Eq.~(\ref{p+_5}) in the Fourier domain:
\begin{equation}
    \left[1+\frac{i\omega}{\mathrm{\Gamma}(1+i\alpha)}\right]{\hat{p}}^+=\mathcal{F}\left[-\left(1+i\alpha\right)\ {(D}_0F^++{D_1}^+ F^-)\right]
\end{equation}
Then, using the Taylor expansion of $(1+x)^{-1}$ truncated at the second order, anti-transforming both sides of the obtained equation, and inserting into Eq.~(\ref{field01}), we have:
\begin{multline}
     \frac{\partial F^+}{\partial\eta}+\frac{\partial F^+}{\partial t^\prime}
    =\sigma\Bigg[\left(\mu-1+i\alpha\mu\right) F^{+}\\-2\left(1+i\alpha \right)\left(|F^+|^2+2|F^-|^2\right)F^++\left(\frac{1}{\Gamma^2\left(1+i\alpha\right)}\right)\frac{\partial^2 F^+}{\partial \eta^2}\Bigg]
\label{modrid1}\end{multline}
%Following the same mathematical treatment also for the backward field, we obtain:
%\begin{multline}
%   -  \frac{\partial F^-}{\partial\eta}+\frac{\partial F^-}{\partial t^\prime}
%    =\sigma\Bigg[\left(\mu-1+i\alpha\mu\right) F^{-}\\-2\left(1+i\alpha \right)\left(|F^-|^2+2|F^+|^2\right)F^-+\left(\frac{1}{\Gamma^2\left(1+i\alpha\right)}\right)\frac{\partial^2 F^+}{\partial \eta^2}\Bigg] \label{modrid2}\end{multline}
%Eqs.~(\ref{modrid1})-(\ref{modrid2}) are the reduced model for the FP configuration, with boundary conditions Eqs.~(\ref{bcc1})-(\ref{bcc2}).

In order to investigate the role in the comb formation of the coupling between $F^+$ and $F^-$ due to the SHB, we introduce into Eqs.~(\ref{modrid1}) the coupling coefficient $K$, so that Eq.~(\ref{modrid1}) becomes:
\begin{align}
     \frac{\partial F^+}{\partial\eta}+\frac{\partial F^+}{\partial t^\prime}
    &=\sigma\Bigl[\left(\mu-1+i\alpha\mu\right) F^{+}\nonumber\\
    &-2\left(1+i\alpha \right)\left(|F^+|^2+2K|F^-|^2\right)F^+\nonumber\\
    &+\left(\frac{1}{\Gamma^2\left(1+i\alpha\right)}\right)\frac{\partial^2 F^+}{\partial \eta^2}\Bigr]\label{modridK1}
    \end{align}
    Following the same mathematical treatment also for the backward field, we obtain:
    \begin{align}
    -\frac{\partial F^-}{\partial\eta}+\frac{\partial F^-}{\partial t^\prime}
   &=\sigma\Bigl[\left(\mu-1+i\alpha\mu\right) F^{-}\nonumber\\
   &-2\left(1+i\alpha \right)\left(|F^-|^2+2K|F^+|^2\right)F^-\nonumber\\
   &+\left(\frac{1}{\Gamma^2\left(1+i\alpha\right)}\right)\frac{\partial^2 F^-}{\partial \eta^2}\Bigr] \label{modridK2}
\end{align}
Eqs.~(\ref{modridK1})-(\ref{modridK2}) are the reduced model for the FP configuration, with boundary conditions Eqs.~(\ref{bcc1})-(\ref{bcc2}). The case $K=1$ corresponds to the FP configuration, while for $K=0$ (no coupling between $F^+$ and $F^-$) we have the unidirectional ring configuration if $R=1$.  For further convenience we define the normalized pump parameter $p=\frac{\mu}{\mu_{\textrm{thr}}}$, where $\mu_{\textrm{thr}}$ is the threshold value of $\mu$.\\
\subsection{Single Complex Ginzburg-Landau equation for the ring configuration}
We can derive the dynamics of the unidirectional ring configuration as a special case of the FP configuration. In fact, by setting $F^-=0$ in Eq.~(\ref{modridK1}) and renaming $F=F^+$, we obtain:
\begin{align}
   \frac{\partial F}{\partial\eta}+\frac{\partial F}{\partial t^{\prime}}
   &=\sigma\Bigl[\left(\mu-1+i\alpha\mu\right) F-2\left(1+i\alpha \right)|F|^2F\nonumber\\
   &+\left(\frac{1}{\Gamma^2\left(1+i\alpha\right)}\right)\frac{\partial^2 F}{\partial \eta^2}\Bigr]
\label{modridring}
\end{align}
We observe that Eq.~(\ref{modridring}) takes the form of a single CGLE for the unidirectional field $F$, in agreement with previous studies on ring QCLs \cite{NaturePiccardo,NB2024,Columbo2021}. Eq.~(\ref{modridring}) is completed by the boundary condition for the ring cavity:
\begin{eqnarray}
F(0,t^\prime)&=&\sqrt{R}F(L^\prime,t^\prime),\label{bcring}
\end{eqnarray}

\section{Single spatiotemporal equation for QCL multimode dynamics}\label{secsingleeq}
\subsection{Derivation}\label{secderivsingleeq}
We can further reduce the two coupled CGLEs for the FP configuration, Eqs.~(\ref{modridK1})-(\ref{modridK2}), to a single spatiotemporal equation for the dynamics of an auxiliary field. The approach used for this derivation is analogous to that followed in \cite{Cole2018} for Kerr frequency combs in FP microresonators. We consider the low transmission limit, assuming $R=1$, and introduce the following modal expansions for the fields $F^+$ and $F^-$ in terms of modal amplitudes $f^{\prime}_n$:
\begin{eqnarray}
   F^+\left(\eta,t^\prime\right)&=&\sum_{n=-\infty}^{+\infty}{f_n^\prime\left(t^\prime\right)e^{i\alpha_n\eta}}\label{svil1}\\
   F^-\left(\eta,t^\prime\right)&=&\sum_{n=-\infty}^{+\infty}{f_n^\prime\left(t^\prime\right)e^{-i\alpha_n\eta}}\label{svil2}
\end{eqnarray}
where $\alpha_n=n \pi/L^{\prime}$.
\\By exploiting Eqs.~(\ref{svil1})-(\ref{svil2}), we extend the domain of definition of $F^+$ and $F^-$ to the interval $\eta \in [-L^\prime ; L^\prime]$ by applying these definitions in the interval $[-L^\prime;0]$:

\begin{eqnarray}
F^+(\eta,t^\prime)=F^-(-\eta,t^\prime)\label{def1}\\
F^-(\eta,t^\prime)=F^+(-\eta,t^\prime)\label{def2}
\end{eqnarray}
Therefore, the forward and backward fields satisfy periodic boundary conditions in the interval $[-L^\prime ; L^\prime]$.\\
We can obtain the modal amplitudes using:
\begin{eqnarray}
     f_n^\prime\left(t^\prime\right)&=&\frac{1}{2L^\prime}\int_{-L^\prime}^{L^\prime}{d\eta e^{-i\alpha_n\eta}F^+(\eta,t^\prime)}\nonumber \\&=& \frac{1}{2L^\prime}\int_{-L^\prime}^{L^\prime}{d\eta e^{i\alpha_n\eta}F^-(\eta,t^\prime)}     \label{fn}                        
\end{eqnarray}
If we use Eqs.~(\ref{svil1})-(\ref{svil2})-(\ref{fn}), Eq.~(\ref{modrid1}) becomes:
\begin{align}
    &\frac{d{f^\prime_n}}{dt^\prime}+i\alpha_n{f^\prime_n}=\sigma\Bigg[\left(\mu-1+i\alpha\mu\right) {f^\prime_n}\nonumber\\
   &-2\left(1+i\alpha \right)\sum_{n^\prime, n^{\prime\prime}}{{f^\prime_{n^{\prime}}}{f^{\prime*}_{n^{\prime\prime}}}\left( {f^\prime_{n-n^\prime+n^{\prime\prime}}}+2K{f^\prime_{-n+n^\prime+n^{\prime\prime}}} \right)}\nonumber\\
 &+\left(\frac{{-\alpha^2}_n}{\Gamma^2\left(1+i\alpha\right)}\right){f^\prime_n}\Bigg]\label{LLE1}
\end{align}
We now introduce the modal amplitudes $f_n$ related to the previously defined $f^\prime_n$ by:
\begin{equation}
    f_n={f^\prime_n} e^{+i\alpha_n t^\prime} \label{fprime}
\end{equation}
%\begin{align}
% &\frac{d{f}_n}{dt^\prime}=\sigma\Bigl[\left(\mu-1+i\alpha\mu\right) {f^\prime}_n\nonumber\\
% &-2\left(1+i\alpha \right)\sum_{n^\prime, n^{\prime\prime}}{{f}_{n^{\prime}}{f^{*}}_{n^{\prime\prime}}\Bigl( {f}_{n-n^\prime+n^{\prime\prime}} 
%+2K{f}_{-n+n^\prime+n^{\prime\prime}}e^{2i\left(\alpha_n-\alpha_{n^\prime}t^\prime\right)} \Bigr)}\nonumber\\
% &+\left(\frac{{-\alpha^2}_n}{\Gamma^2\left(1+i\alpha\right)}\right){f}_n\Bigr]\label{LLE2}
%\end{align}
Inserting Eq.~(\ref{fprime}) in Eq.~(\ref{LLE1}) and averaging the resulting equation over a time interval longer than the cavity round trip and shorter than the cavity decay time, we obtain:
\begin{align}
      &\frac{d{f}_n}{dt^\prime} =\sigma\Bigg[\left(\mu-1+i\alpha\mu\right) f^\prime_n\nonumber\\ &-2\left(1+i\alpha \right)\left(\sum_{n^\prime, n^{\prime\prime}}{{f}_{n^{\prime}}{f^{*}_{n^{\prime\prime}}}{f}_{n-n^\prime+n^{\prime\prime}}} +2K\sum_{n^\prime}{{f}_{n^{\prime}}{f^{*}_{n^{\prime}}}}\right)\nonumber\\
      &+\left(\frac{{-\alpha^2}_n}{\Gamma^2\left(1+i\alpha\right)}\right){f}_n\Bigg]\label{LLE3}
\end{align}
We now define the auxiliary field:
\begin{eqnarray}
\psi\left(\eta,t^\prime\right)&=&\sum_{n=-\infty}^{+\infty}{f_n\left(t^\prime\right)e^{i\alpha_n\eta}}\label{psi_def}
\end{eqnarray}
Then, Eq.~(\ref{LLE3}) becomes:
\begin{align}
&\frac{\partial \psi}{\partial t^{\prime}}
    =\sigma\Bigg[\left(\mu-1+i\alpha\mu\right) \psi-2\left(1+i\alpha \right)|\psi|^2\psi\nonumber\\
    &-4K\left(1+i\alpha\right)\psi\frac{1}{2L^{\prime}}\int_{-L^\prime}^{L^\prime}{d\eta|\psi|^2}+\left(\frac{1}{\Gamma^2\left(1+i\alpha\right)}\right)\frac{\partial^2 \psi}{\partial \eta^2}\Bigg]\label{psieq}
\end{align}
Eq.~(\ref{psieq}) is a single spatiotemporal equation with periodic boundary conditions for the auxiliary field $\psi$, which serves as the order parameter, and is equivalent to Eqs.~(\ref{modridK1})-(\ref{modridK2}) when $R=1$. Given this equivalence, if the spatiotemporal evolution of $\psi$ is known, it is possible to retrieve the dynamics of $F^+$ and $F^-$ using an appropriate reconstruction procedure. Firstly, from the definition of the field $\psi$ Eq.~(\ref{psi_def}), we can retrieve the coefficient $f_n$:
\begin{eqnarray}
     f_n&=&\frac{1}{2L^\prime}\int_{-L^\prime}^{L^\prime}{d\eta e^{-i\alpha_n\eta}\psi(\eta,t^\prime)} \label{f_n}        
\end{eqnarray}
Then, inverting Eq.~(\ref{fprime}), we can retrieve $f^\prime_n$ from the coefficients $f_n$:
\begin{equation}
    {f^\prime_n}=f_n e^{-i\alpha_n t^\prime} \label{f_n_from_f_prime}
\end{equation}
Finally, we can replace the expression obtained for $f^{\prime}_n$ into Eq.~(\ref{svil1}), to obtain the field $F^+$; by replacing  $f^{\prime}_n$ into Eq.~(\ref{svil2}) we retrieve $F^-$. A numerical test on the equivalence between the two approaches is provided in the Supplementary Material, where for a given set of parameters, the same comb regime was reproduced with good agreement using the two different approaches.

We highlight that a single spatiotemporal equation for FP QCL dynamics was also reported in \cite{Burghoff20}, albeit with a notable distinction compared to our theory. In fact, in \cite{Burghoff20} a single equation was derived by eliminating the dynamics of the field amplitude and assuming a constant amplitude. Although this method enables one to correctly reproduce the phase (or instantaneous frequency) dynamics, it cannot allow for replicating the simultaneous presence of AM and FM behaviors observed in QCL combs. The AM features in QCL combs are in fact significant and non-negligible, as highlighted by various experiments utilizing different techniques such as SWIFTS \cite{Optica_Faist_18}, FACE \cite{Cappelli2019}, and electro-optic sampling combined with computational phase correction \cite{Faist_comb_profile_2023}.
In contrast, our theory retains the dynamics of the field amplitude and, as presented in Section~\ref{sec_num}, our model successfully reproduces the typical amplitude modulations characteristic of QCL combs, along with the modulation of the instantaneous frequency.

Furthermore, we emphasize that Eq.~(\ref{psieq}) takes the form of a CGLE modified by the presence of a non-local integral term proportional to the coupling coefficient $K$. This term accounts for the coupling between the counterpropagating fields $F^+$ and $F^-$ in the laser cavity due to SHB. The implications of this term on the formation of combs are discussed in Sect.~\ref{secroleK}.

A final note concerns the definitions~(\ref{def1})-(\ref{def2}) that allowed for extending the equations~(\ref{modridK1})-(\ref{modridK2}) to the interval $[-L^\prime,L^\prime]$. This enabled the use of periodic boundary conditions for the fields $F^+$ and $F^-$, thereby allowing the utilization of traveling waves rather than standing waves, leading to a significant simplification in the derivation \cite{Cole2018}.
\begin{figure*} [t]
   \begin{center}
%   \begin{tabular}{c} %% tabular useful for creating an array of images 
   \includegraphics[width=1\textwidth]{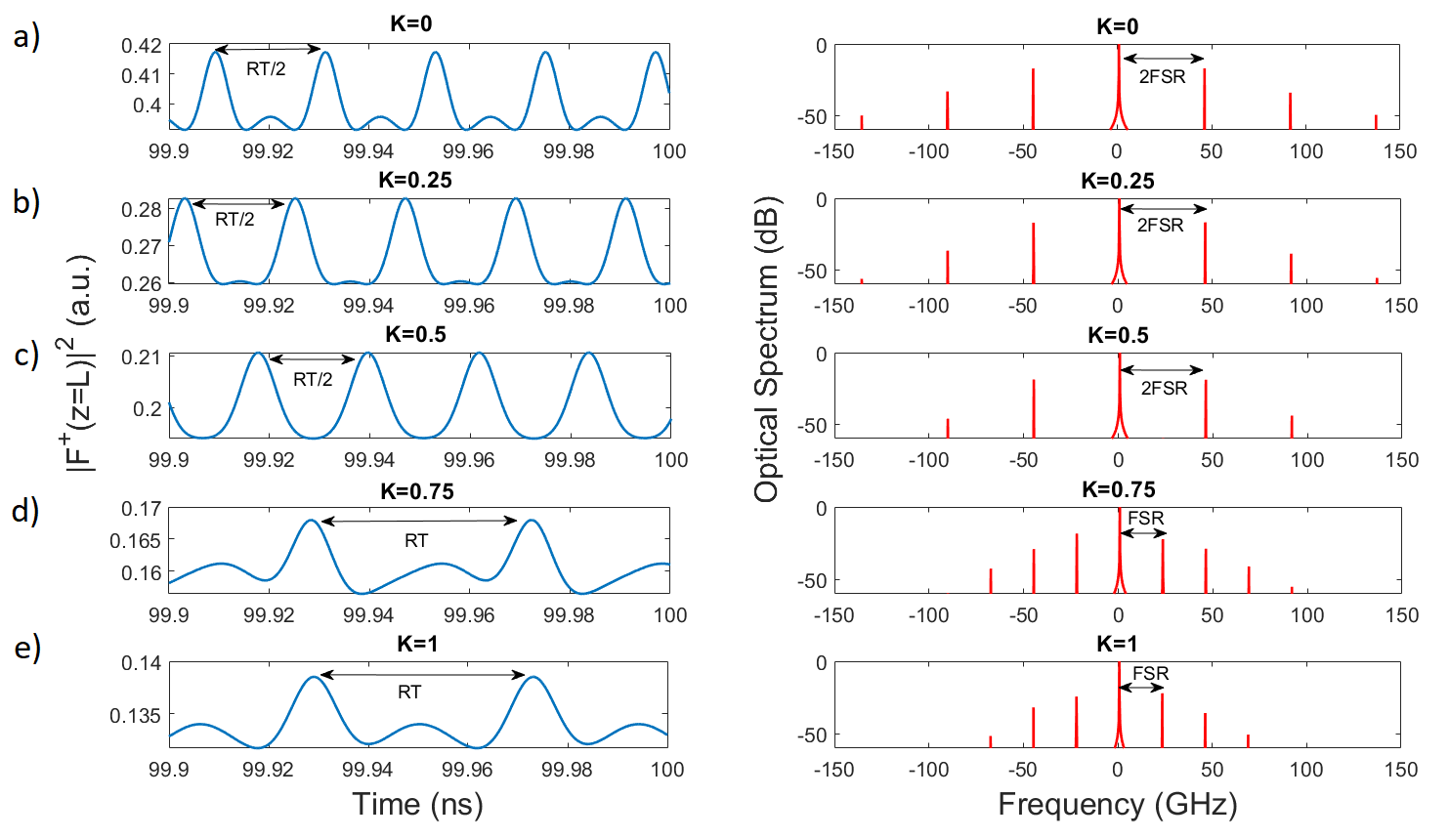}
%   \end{tabular}
   \end{center}
   \caption{The impact of the nonlinear integral term in Eq.~(\ref{psieq}) on the formation of multiple structures per round trip, associated with harmonic comb emission. Normalized power as a function of time (left) and optical spectrum (right) for different regimes obtained by integrating Eq.~(\ref{psieq}) with $K=0$ (a), $K=0.25$ (b), $K=0.5$ (c), $K=0.75$ (d) and $K=1$ (e). The other parameters are $\alpha=1.15$, $\tau_d=0.1~\mathrm{ps}$, $\Gamma=0.06$, $p=1.8$, $\sigma=4.5 \times 10^{-4}$, and $L=2~\mathrm{mm}$. Second-order HFC emission is reported for $K=0, 0.25, 0.5$, while fundamental OFCs form for $K=0.75$ and $K=1$. The case $K=0$ corresponds to the unidirectional ring cavity, while $K=1$ reproduces the FP configuration.} \label{fig1}
\end{figure*}
\subsection{Role of the non-local integral term}\label{secroleK}
%In Sec.~\ref{secderivsingleeq} we derived Eq.~(\ref{psieq}), i.e. a single spatiotemporal equation for the order parameter $\psi$ describing the multimode dynamics of FP QCLs. We highlighted the presence of a non local integral term proportional to the coupling coefficient K, which accounts for the coupling between the counterpropagating fields in the FP cavity. We discuss here the impact of this term on the comb formation.\\
We discuss here the impact of the non-local integral term appearing in Eq.~(\ref{psieq}) on the comb formation.\\
We begin by observing that Eq.~(\ref{psieq}) can be configured for the ring case by suppressing this non-local integral term, i.e., by setting $K=0$. In that case, $\psi$ reduces to the field $F$, and 
Eq.~(\ref{psieq}) becomes Eq.~(\ref{modridring}).\\Therefore, our formalism unifies in a single equation both ring and FP QCL dynamics, which can be obtained by suppressing (ring) and keeping (FP) the non-local integral term. For this reason, Eq.~(\ref{psieq}) offers the opportunity to study the effect of the coupling between the fields on the formation of the localized structures, allowing us to establish certain differences between the two configurations.

To gain insight into this aspect, we solve Eq.~(\ref{psieq}) in the parameter set $\alpha=1.15$, $\tau_d=0.1~\mathrm{ps}$, $\Gamma=0.06$, $p=1.8$, $\sigma=4.5 \times 10^{-4}$, and with a cavity length of 2 mm. We remark that the considered value of $\alpha$ is in agreement with experimental measures reported in the literature for this parameter (see, e.g., \cite{Grillot16}), while $\Gamma=0.06$ corresponds to a gain curve width of about 200 GHz, consistent with the values of this quantity for single-stack THz QCLs \cite{Vitiellorev2}.\\
Firstly, we solve Eq.~(\ref{psieq}) for $K=0$, i.e. in the unidirectional ring case. In Fig.~\ref{fig1}(a), the reconstructed intensity $|F^+|^2$ is plotted as a function of time (left) along with the corresponding optical spectrum (right). Note that in this case $F^+$ corresponds to the unidirectional field $F$ appearing in Eq.~(\ref{modridring}). We observe a regular repetition of field structures with a period half the cavity round-trip time (RT/2), which corresponds to a comb spectrum with spacing twice the free-spectral range (2~FSR) of the QCL cavity. Therefore, a second-order harmonic comb is reported for $K=0$. If we increase $K$ to 0.25 and 0.5, we still observe second-order HFCs, as depicted, respectively, in Figs.~\ref{fig1}(b) and \ref{fig1}(c). However, if the coupling strength is further increased by setting $K=0.75$, we report a transition to a dense comb regime, i.e. a comb with spacing corresponding to the cavity FSR (see Fig.~\ref{fig1}(d)), which is also observed in the pure FP case $K=1$, as presented in Fig.~\ref{fig1}(e).

These results show how an increase in the field coupling impacts the formation of structures. In particular, the multiple structures per round trip observed in the ring case, associated with the formation of harmonic combs, disappear when $K$ reaches a certain magnitude and are replaced by fundamental combs. In this sense, the results presented in Fig.~\ref{fig1} suggest an anti-patterning role of the nonlinear integral term, implying a greater predisposition of the ring configuration to form harmonic combs compared to the FP.
\begin{figure*} [t!]
   \begin{center}
%   \begin{tabular}{c} %% tabular useful for creating an array of images 
   \includegraphics[width=1\textwidth]{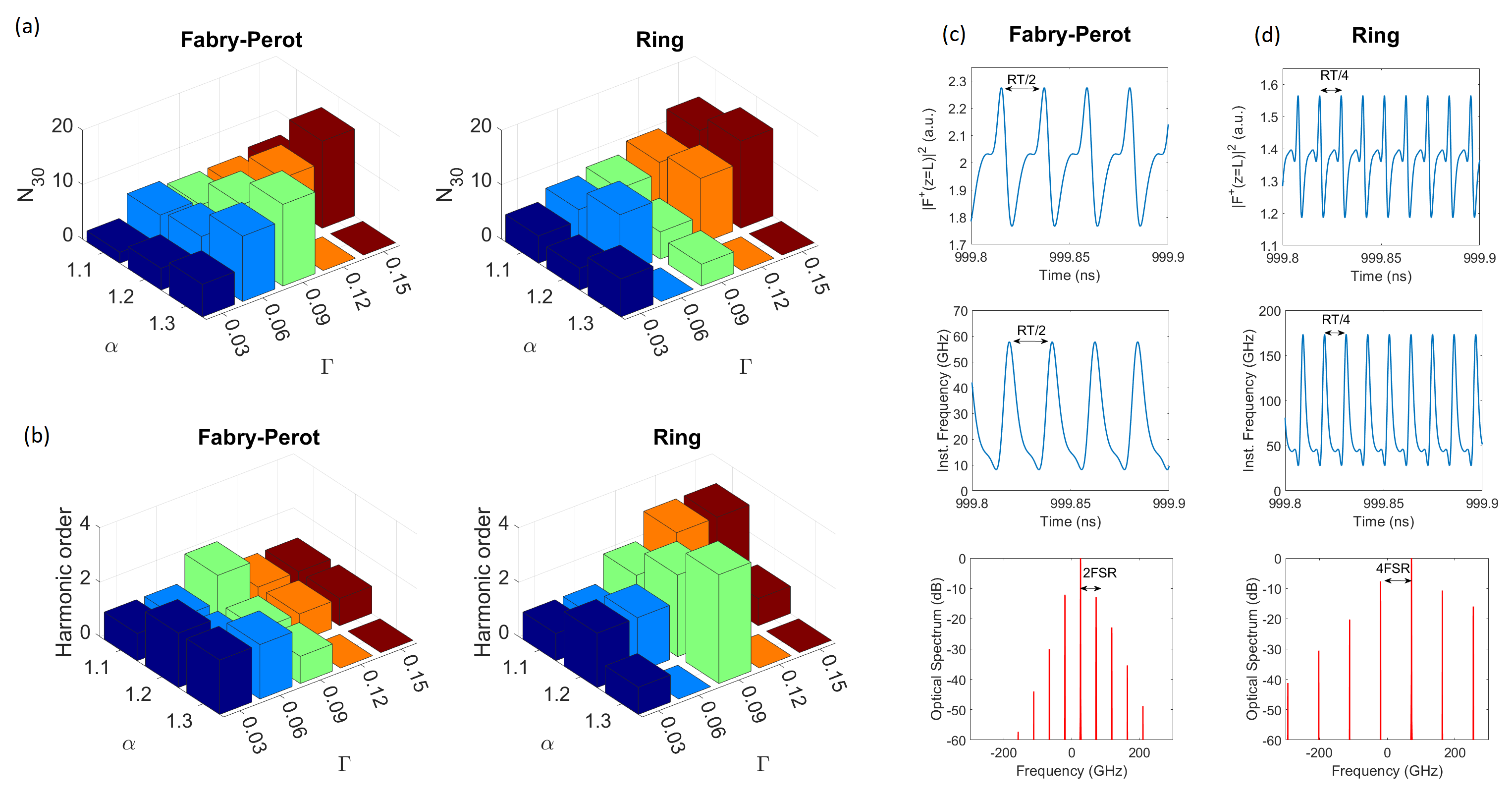}
%   \end{tabular}
   \end{center}
   \caption{(a)-(b) Histograms showing respectively the maximum number of locked optical modes in the $-30~$dB band $N_{30}$ and the maximum comb harmonic order reported for each pair ($\alpha$, $\Gamma$), for both FP (left) and ring (right) configurations; $\alpha\in\left[1.1, 1.3\right]$ and $\Gamma\in\left[0.03, 0.15\right]$. The normalized pump parameter $p$ is swept between 1.1 and 2 for each pair ($\alpha$, $\Gamma$). The other parameters are $\sigma=4.5\times 10^{-4}$, $\tau_d=0.1~$ps. The cavity length is $L=$2 mm for the FP and $L=$4 mm for the ring in order to have the same FSR. (c)-(d) Harmonic combs reported in the FP and ring cases. In (c), a second-order FP HFC is shown, with its temporal evolution of normalized power (top) and instantaneous frequency (center), and its optical spectrum (bottom). In (d), the same quantities are plotted for a fourth-order ring HFC. } \label{fig2}
\end{figure*}
\section{Numerical results}\label{sec_num}
\subsection{Comparative analysis of FP and ring configurations}\label{sec_compar}
One of the key advantages of the reduced models obtained in Sections \ref{Sec_deriv} and \ref{secsingleeq} is that they allow for systematic sets of simulations. This enables reproducing dynamic regimes and scenarios compatible with experiments and characterizing the obtained comb states while varying different QCL parameters, such as the gain curve width, the $\alpha$ factor, and the pump.\\For this purpose, we adopt the reduced models presented in Section \ref{Sec_deriv}, as they allow us to relax the assumption of periodic boundary conditions, enabling us to use reflectivity values $R$ close to those reported in experiments. Therefore, we set $R=0.3$ for all simulations presented in this section.\\
Regarding the FP configuration, we numerically solve the two coupled CGLEs Eqs.~(\ref{modridK1})-(\ref{modridK2}) with boundary conditions Eqs.~(\ref{bcc1})-(\ref{bcc2}) with $K=1$, $\sigma=4.5\times 10^{-4}$, $\tau_d=100~$fs, $L=$2~ mm, for different pairs ($\alpha$, $\Gamma$), so that $\alpha\in\left[1.1, 1.3\right]$ and $\Gamma\in\left[0.03, 0.15\right]$. The chosen values for $\alpha$ agree with the experimental measurements reported in \cite{Grillot16}, while the selected range for $\Gamma$ corresponds to gain bandwidth values between 100 GHz and 500 GHz, typical for THz QCLs based on a single-stack active region \cite{Vitiellorev2}. For each pair ($\alpha$, $\Gamma$), we perform a scan of the normalized pump parameter $p$ between 1.1 and 2, with an increment of 0.1. Then, we replicate these simulations for the ring configuration by solving the single CGLE~(\ref{modridring}) with boundary condition Eq.~(\ref{bcring}), using the same parameters as in the FP case. This allows for a comparison between the two configurations with respect to the characteristics of the reported OFCs. We remark that to maintain the same free spectral range in both schemes, we assumed for the ring a cavity length twice that of the FP. To characterize the comb states that we found, we used the maximum number of locked modes within the $-30~$dB spectral bandwidth from the peak of the optical spectrum, denoted as $N_{30}$, and the maximum harmonic order of frequency combs, reported for each pair ($\alpha$, $\Gamma$). The results are summarized in Figs.~\ref{fig2}(a)-(b), where the values of the two figures of merit are plotted as a function of $\alpha$ and $\Gamma$ for both configurations.\\
\begin{figure} [t]
   \begin{center}
%   \begin{tabular}{c} %% tabular useful for creating an array of images 
   \includegraphics[width=0.49\textwidth]{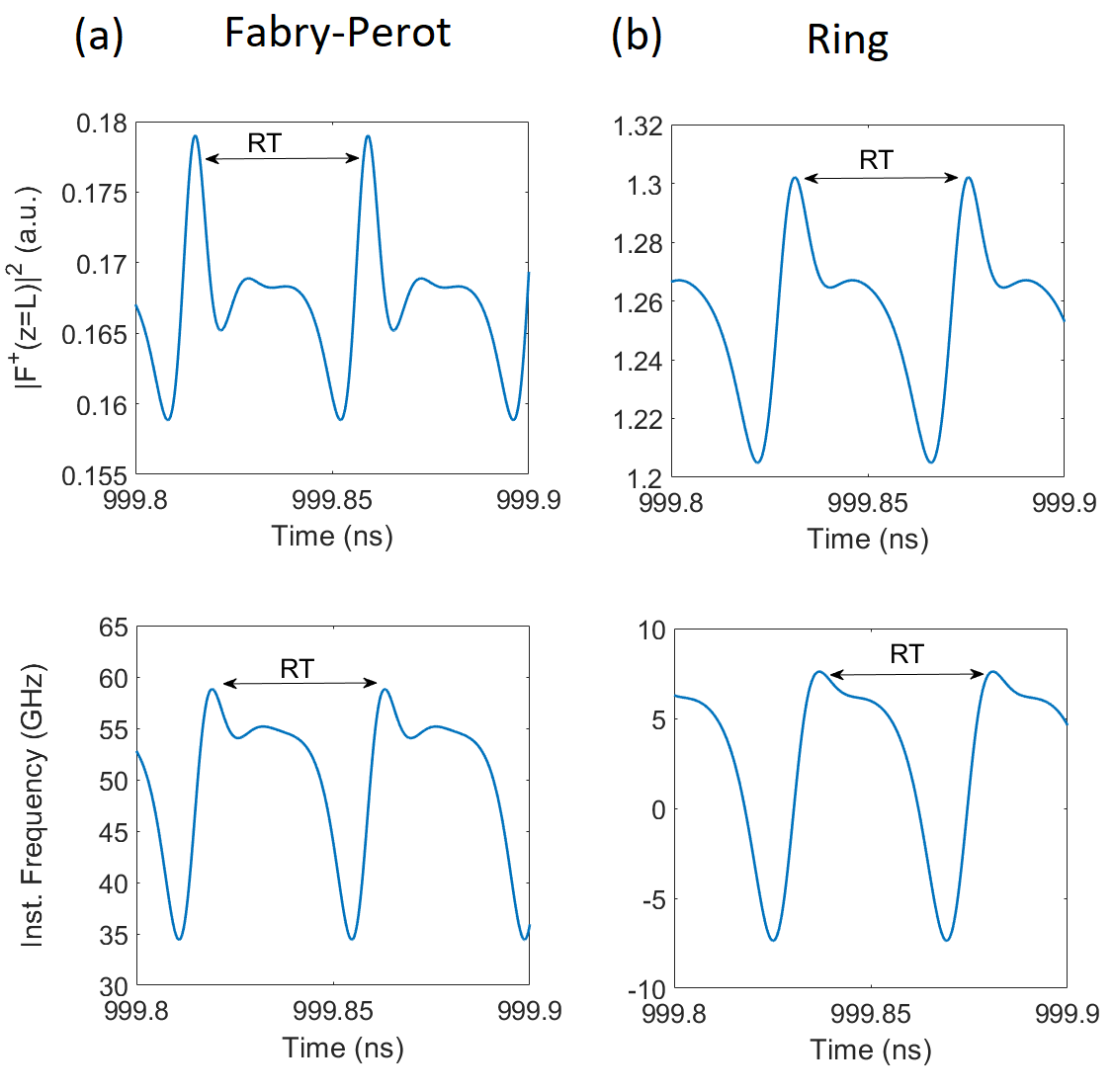}
%   \end{tabular}
   \end{center}
   \caption{Example of simulated fundamental combs in FP (a) and ring (a) QCLs: normalized power (top) and instantaneous frequency (bottom) as a function of time. The FP comb has been obtained for $\alpha=1.2$, $\Gamma=0.09$, $p=1.1$. The ring comb corresponds to $\alpha=1.1$, $\Gamma=0.03$, $p=1.5$. Both regimes have been extracted from the simulations presented in Figs.~\ref{fig2}(a)-(b). } \label{figOFC}
\end{figure}
Upon examination of Fig.~\ref{fig2}(a), it should be noted that the maximum number of comb lines $N_{30}$ remains consistent across both configurations, namely 16. Furthermore, this value is attained for the same pair $\alpha=1.2$, $\Gamma=0.15$.  However, we can notice some differences in the distribution of the number of modes between the two histograms Fig.~\ref{fig2}(a). For example, for $\alpha = 1.2$, in the FP configuration we observe a steady increase with $\Gamma$. However, in the case of the ring resonator, $N_{30}$ decreases from 11 to 5 as $\Gamma$ varies from 0.06 to 0.09. Subsequently, it attains the value $N_{30} = 12$ for $\Gamma = 0.12$. This phenomenon arises due to the occurrence of only harmonic combs for $\Gamma = 0.09$, without any fundamental combs being present, thereby resulting in a reduction in the number of modes. Indeed, in harmonic combs, since some cavity modes are suppressed, the number of optical lines is lower compared to fundamental combs, for the same spectral width. Regarding the formation of harmonic combs, in fact, the two configurations exhibit more pronounced differences, as can be observed in Fig.~\ref{fig2}(b). Overall, we document a higher number of harmonic states within the unidirectional ring cavity compared to the FP configuration. Moreover, the ring setup demonstrates the ability to generate comb regimes with increased harmonic orders. Specifically, we report a maximum harmonic comb order of 2 in the FP case, whereas in the ring configuration, HFCs of order 3 and 4 can also be observed. An example of second-order HFC within the FP configuration, depicted by the temporal evolution of its power/instantaneous frequency and optical spectrum, is shown in Fig.~\ref{fig1}(c). This harmonic comb is characterized by a coexistence of AM and FM features, exhibiting a regular repetition of both power and instantaneous frequency structures, with a period equal to half of the round trip time of the laser cavity. Its dual representation in the frequency domain corresponds to an optical spectrum with equally spaced lines at twice the FSR. As mentioned, in the ring case we also observe fourth-order HFCs, with a temporal period of RT/4 and an optical line spacing equal to 4 times the FSR (see Fig.~\ref{fig1}(d)).
The numerical results presented in Fig.~\ref{fig2} are consistent with the analysis conducted in Section~\ref{secroleK} on the anti-patterning role of the non-local integral term in Eq.~(\ref{psieq}). Indeed, we have observed how the coupling between forward and backward fields, characteristic of the FP configuration and absent in the ring case, is detrimental to the formation of multiple localized structures, which manifest themselves in the form of harmonic combs. This could explain both the lower occurrence of HFCs and the lower harmonic order in the FP case, reported in the simulations summarized in Fig.~\ref{fig2}(b).\\
%%FUNDAMENTAL COMB
Regarding the simulated fundamental combs, An illustrative example within the FP configuration is showcased in Fig.~\ref{figOFC}(a). We observe a hybrid AM-FM behavior in the temporal profiles of power (top panel) and instantaneous frequency (bottom panel), consistent with experimental findings \cite{Optica_Faist_18, Cappelli2019} and with the numerical results obtained with the full model ESMBEs \cite{Silvestri20,Silvestri22}. Specifically, the power structures exhibit a recurring pattern with each round-trip (top panel), characterized by a primary AM peak and two secondary bumps. Furthermore, the trace of instantaneous frequency reveals a sequence of features deviating from a linear chirp, comprising a primary peak and a quasi-flat region. Consequently, this comb exhibits qualitative characteristics consistent with experimentally reported waveforms for THz-QCL OFCs (see Fig.~5 in \cite{Cappelli2019}). This alignment is consistent with the used value of $\Gamma$, which corresponds to a gain bandwidth of 300 GHz, typical for THz QCLs \cite{Vitiellorev2}. %Furthermore, also the number of modes in optical spectrum (bottom panel in Fig.~\ref{figOFC}(a)) is consistent with experiments \cite{Cappelli2019}.
On the other hand, the temporal traces for a ring comb presented in Fig.~\ref{figOFC}(b) are also consistent with those obtained in experiments using the SWIFTS technique \cite{Bomeng1}, and closely resemble the ones reproduced with the ring ESMBEs (e.g., see Fig.~6(a) in \cite{Columbo2018}).

\begin{figure} [t!]
   \begin{center}
%   \begin{tabular}{c} %% tabular useful for creating an array of images 
   \includegraphics[width=0.5\textwidth]{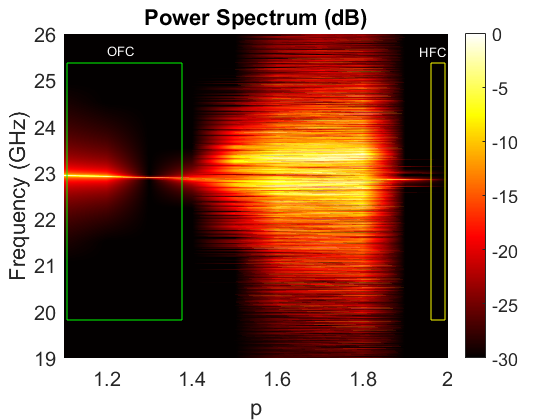}
%   \end{tabular}
   \end{center}
   \caption{First beatnote observed in the power spectrum while varying the pump parameter $p$ in the FP configuration, with $\Gamma=0.03$, $\alpha=1.3$, and other parameters as specified in Fig.~\ref{fig2}. We highlight in green the region of dense OFCs, and in yellow the region of HFCs.} \label{figRF}
\end{figure}
%%RF MAP
Furthermore, we highlight that these reduced models allow us to reproduce dynamic scenarios sweeping the pump, consistent with experimental evidence. An example is shown in Fig.~\ref{figRF}, where we present a power spectrum map with a zoom on the first beatnote, corresponding to one of the pairs ($\alpha$, $\Gamma$) of Fig.~\ref{fig2}. Here, we report the formation of fundamental OFCs, unlocked states characterized by a broad beatnote, and harmonic regimes where the first beatnote is notably absent. This scenario aligns with the experimental findings described, for example, in \cite{Li15}. Notably, it also qualitatively concurs with the full model ESMBEs. For instance, Fig.~5 in \cite{Silvestri22}  illustrates a beatnote map for a FP THz-QCL similar to Fig.~\ref{figRF}(c).
%LINEAR CHIRP
\subsection{Observation of linear chirp\\for large gain bandwidth}
Increasing $\Gamma$ reduces the effective polarization dephasing time, necessitating a corresponding decrease in the simulation time step. This would significantly prolong simulation durations, and make impractical within a reasonable timeframe the possibility to run  hundreds of simulations. Therefore, for numerical reasons, the systematic simulations presented in Sec.~\ref{sec_compar} were conducted for $\Gamma$ values ranging from 0.03 to 0.15. However, we have examined comb properties also for some higher values of $\Gamma$.
\begin{figure} [t!]
   \begin{center}
%   \begin{tabular}{c} %% tabular useful for creating an array of images 
   \includegraphics[width=0.45\textwidth]{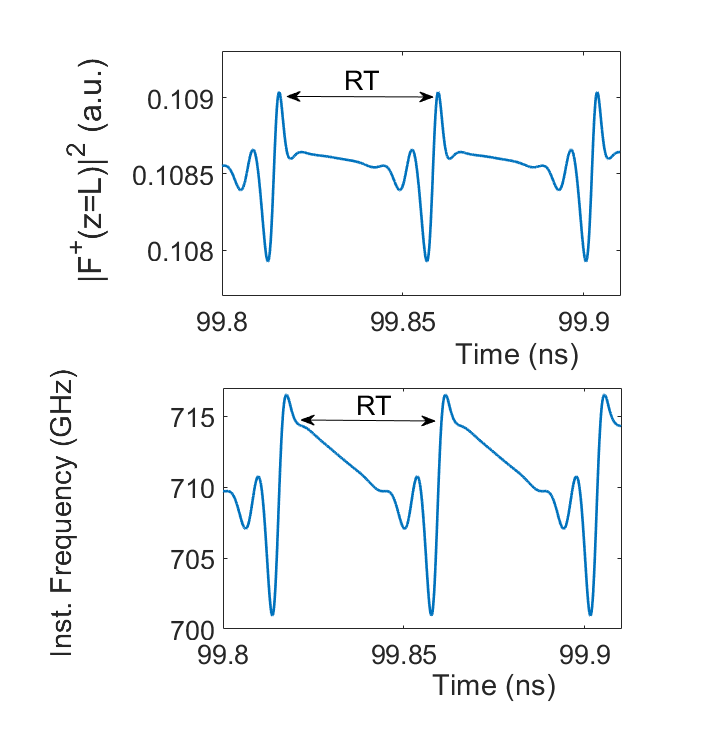}
%   \end{tabular}
   \end{center}
   \caption{Simulated OFC in FP QCLs for $\Gamma=0.5$: normalized power (top) and instantaneous frequency (bottom) as a function of time. The other parameters are as in Fig.~\ref{figOFC}(a). Typical features of mid-IR combs are observable, such as power spikes on a constant background and linear chirp of the instantaneous frequency. } \label{figchirp}
\end{figure}
An example of an FP comb for $\Gamma=0.5$ is depicted in Fig.~\ref{figchirp}. It is noteworthy that the power trace (top panel) bears resemblance to the one obtained with the full model ESMBEs (see Fig. 2 in \cite{Silvestri20}), characterized by power spikes resting on an almost constant background. Furthermore, the instantaneous frequency (bottom panel) also exhibits the typical linear chirp trend characteristic of the mid-IR region (see Fig. 5 in \cite{Optica_Faist_18}), which was also reproduced in \cite{Silvestri20} with ESMBEs. Hence, we observe agreement with both experiments and the full model on mid-IR combs. This appears reasonable because mid-IR QCLs are characterized by gain bandwidth values higher than 1 THz \cite{Faist_2016}, and the utilized value $\Gamma=0.5$ corresponds to a gain bandwidth of 1.6 THz. We also emphasize that all parameter values except $\Gamma$ remain the same as in Fig.~\ref{figOFC}(a), facilitating a direct comparison. With $\Gamma=0.09$, we observed temporal traces that resemble the experimental results for THz QCLs, whereas for $\Gamma=0.5$, typical features of mid-IR QCLs are evident. This result suggests that the fundamental difference between mid-IR and THz QCLs, enabling linear chirp observation in the former but not in the latter, lies in the presence of a broader gain curve in devices operating in the mid-IR.\\
Finally, we have also verified that, in the ring case, linear chirp behavior is not observed with the same parameters as in Fig.~\ref{figchirp}. This suggests that as $\Gamma$ increases, a clearer difference in the structure of the combs can be observed between the two configurations.
\section{Linear stability analysis \\for the single-mode solution \\in the ring configuration}\label{sec_LSA}
In Sec.~\ref{sec_num} we explored how the presented reduced models allow for systematic numerical simulations. However, these models offer an additional advantage: due to their lower mathematical complexity with respect to full models, they are more amenable to analytical treatments, such as linear stability analysis (LSA). In this section, we delve into how the LSA of the single-mode solution for the reduced model of ring QCLs, Eq.~(\ref{modridring}), holds significant predictive power, providing further insights regarding aspects addressed in the previous sections such as the emergence of multimode regimes and the formation of harmonic combs.
\subsection{Steady state solutions}\label{secsteady}
As an initial step, we calculate the steady-state solutions of Eq.~(\ref{modridring}). Consider, for the field, the following CW expression:
\begin{eqnarray}
F&=&a_0e^{-iq\eta+i\omega t^{\prime}}
\label{cwfield}
\end{eqnarray}
Here, $a_0$ represents the amplitude of the field, which we can assume real without loss of generality, while $\omega$ and $q$ denote the dimensionless angular frequency and wavenumber, respectively.
Upon substitution of Eq.~(\ref{cwfield}) into Eq.~(\ref{modridring}), we derive:

\begin{align}
-iq+i\omega&=\sigma\Bigg[\Bigl(\mu-1+i\alpha\mu\Bigr)\nonumber\\&-2\left(1+i\alpha \right)|a_0|^2+\left(\frac{1}{\Gamma^2\left(1+i\alpha\right)}\right)\left(-q^2\right)\Bigg]
\label{sspass2}
\end{align}

Taking the real part of Eq.~(\ref{sspass2}) we obtain:
\begin{eqnarray}
\left|a_0\right|^2=\frac{1}{2}\left[\mu-1- q^2\left(\frac{1}{\Gamma^2\left(1+\alpha^2\right)}\right)\right]
\label{ssreal}
\end{eqnarray}
which gives the laser intensity versus the pump and the continuous wave wavevector.\\
Taking the imaginary part of Eq.~(\ref{sspass2}) and replacing $|a_0|^2$ with its expression in Eq.~(\ref{ssreal}) we have:
\begin{equation}\omega=q+\sigma\alpha\left(1+\frac{2q^2}{\mathrm{\Gamma}^2\left(1+\alpha^2\right)}\right)\label{ssim}\end{equation}
Eq.~(\ref{ssim}) is the dispersion relation, i.e. the relation between the wavenumber $q$ and the pulsation $\omega$. Eqs.~(\ref{ssreal})-(\ref{ssim}) represent the steady state solutions of Eq.~(\ref{modridring}). We want to comment on the role of $\alpha$ and $\sigma$. From Eq.~(\ref{ssim}), we can notice that if $\alpha=0$, we have $\omega=q$. Furthermore, since in QCLs $\tau_d\approx100~\mathrm{fs}$, while $\tau_p\approx100~\mathrm{ps}$, $\sigma \ll 1$ so that the second term on the right-hand side of Eq.~(\ref{ssim}) is small or negligible. Additionally, we observe that $\alpha$ appears in the last term on the right-hand side of Eq.~(\ref{ssreal}), indicating that its value influences the lasing threshold for $q \neq 0$."
\subsection{Derivation of the linear stability analysis}
We introduce a perturbation $\delta a(\eta,t^{\prime})$ in the CW field expression:
\begin{equation}
    F=\left(a_0+\delta a(\eta,t^{\prime})\right)e^{-iq\eta+i\omega t^{\prime}}\label{cwfe}
\end{equation}
Substituting Eq.~(\ref{cwfe}) into Eq.~(\ref{modridring}), we obtain:
\begin{align}
    &-iq\left(a_0+\delta a\right)+\frac{\partial\delta a}{\partial\eta}+i\omega\left(a_0+\delta a\right)+\frac{\partial\delta a}{\partial t^{\prime}}\nonumber\\
    &=\sigma \Bigg[
    -\left(a_0+\delta a\right)+\left(1+i\alpha\right)\mu\left(a_0+\delta a\right)\nonumber\\&-2\left(1+i\alpha\right)\left|\left(a_0+\delta a\right)\right|^2\left(a_0+\delta a\right)\nonumber\\&+ \left(\frac{1-i\alpha}{\mathrm{\Gamma}^2\left(1+\alpha^2\right)}\right)\left( \left(-q^2\right)\left(a_0+\delta a\right)-2iq\frac{\partial\delta a}{\partial\eta}+\frac{\partial^2\delta a}{\partial\eta^2} \right) \Bigg]
\end{align}

If we use the steady state solutions Eqs.~(\ref{ssreal})-(\ref{ssim}) and we neglect the terms $O((\delta a)^n)$ with n$\geq$2, we have:

\begin{equation}\begin{split}
    -iq\delta a &+\frac{\partial\delta a}{\partial\eta}+i\omega\delta a+\frac{\partial\delta a}{\partial t^{\prime}}\\&=\sigma \Bigl[
    -\delta a+\left(1+i\alpha\right)\mu\delta a-2\left(1+i\alpha\right)
    \left(
    \left|a_0\right|^2\delta a + a_0^2 \delta a^*\right)\\&+ \left(\frac{1-i\alpha}{\mathrm{\Gamma}^2\left(1+\alpha^2\right)}\right)\left( \left(-q^2\right)\delta a-2iq\frac{\partial\delta a}{\partial\eta}+\frac{\partial^2\delta a}{\partial\eta^2} \right) \Bigr]
    \end{split}\label{lsa1}
\end{equation}
We now assume the following Fourier expansions:
\begin{eqnarray}
\delta a &=&\left(\sum_{n=-\infty}^{+\infty}{{{\delta a}_0}_ne^{-ik_n\eta}}\right)e^{\lambda t}\label{exp1}\\
{\delta a}^\ast &=&\left(\sum_{n=-\infty}^{+\infty}{{{{\delta a}_0}^{\phantom{-}\ast}_{-n}} e^{-ik_n\eta}}\right)e^{\lambda t}\label{exp2}
\end{eqnarray}
Substituting Eqs.~(\ref{exp1})-(\ref{exp2}) into Eq.~(\ref{lsa1}), and using the ortonormality of the Fourier basis, we have:
\begin{align}
&-iq{{\delta a}_0}_n -ik_n{{\delta a}_0}_n+i\omega{{\delta a}_0}_n+\lambda{{\ \delta a}_0}_n\nonumber\\ &=\sigma \Bigl[-{{\delta a}_0}_n+\left(1+i\alpha\right)\mu {{\ \delta a}_0}_n \nonumber\\
&-2\left(1+i\alpha\right)\left( 2\left|a_0\right|^2{{\delta a}_0}_n+{a_0}^2{{{\delta a}_0}_{-n}^{\phantom{-}\ast}} \right)\nonumber\\&+\left(\frac{1-i\alpha}{\mathrm{\Gamma}^2\left(1+\alpha^2\right)}\right) \left(-q^2{{\delta a}_0}_n-2iq\left(-ik_n\right){{\delta a}_0}_n-{k_n}^2{{\delta a}_0}_n\right)\Bigr]\label{lsa2} 
\end{align}

\begin{figure*} [t!]
   \begin{center}
%   \begin{tabular}{c} %% tabular useful for creating an array of images 
   \includegraphics[width=0.9\textwidth]{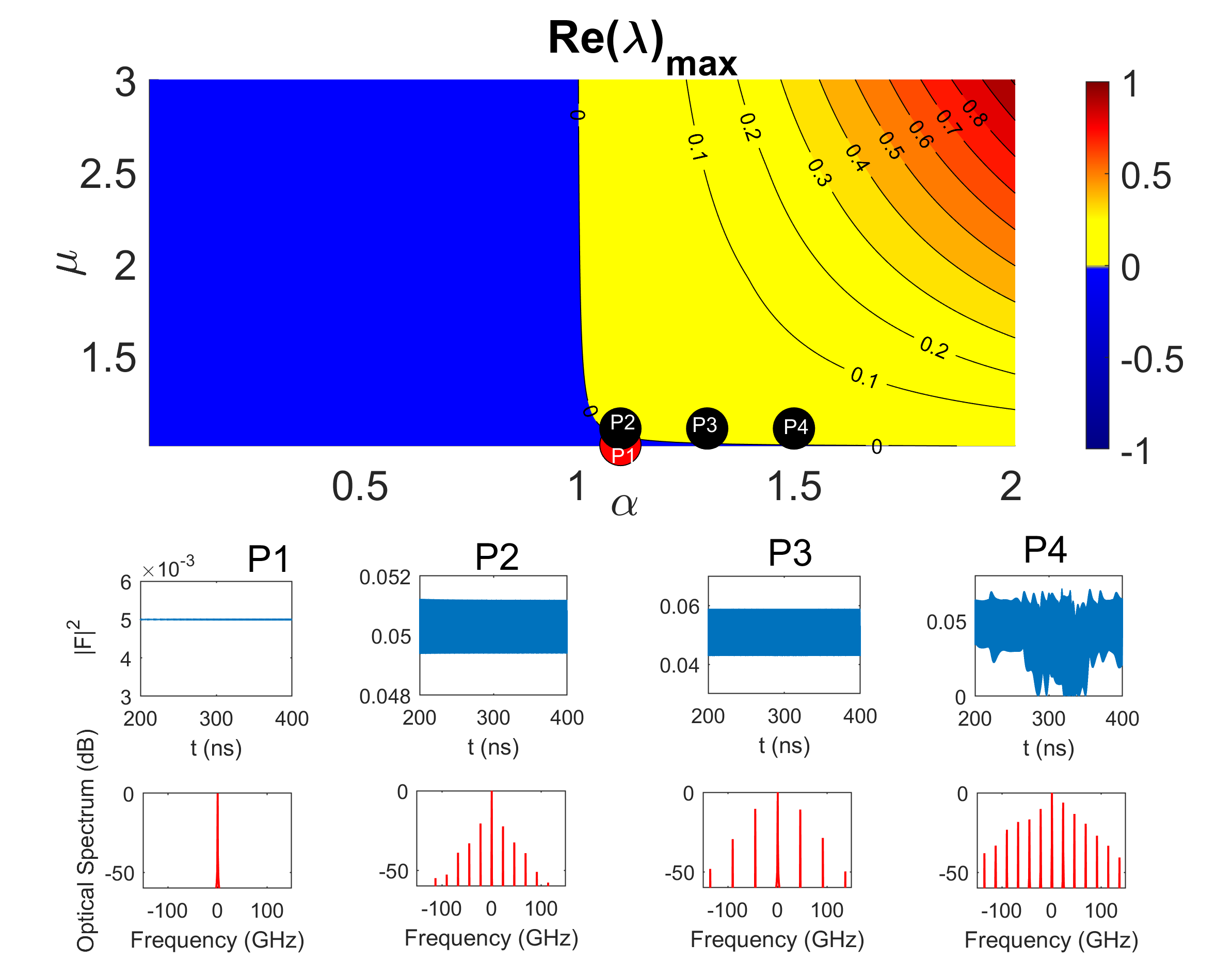}
%   \end{tabular}
   \end{center}
   \caption{Linear stability analysis map for $\Gamma=0.09$, $L=4$ mm, $\sigma=4.5\times 10^{-4}$, $\tau_d=0.1~$ps. The maximum value of $\text{Re}(\lambda)$ is plotted for pairs $(\alpha, \mu)$ with $\alpha$ ranging from 0.015 to 2, and $\mu$ ranging from 1.005 to 3. Negative values of $\text{Re}(\lambda)$, indicative of stable single-mode emission, correspond to the blue region in the map. The red and black dots represent the numerical simulations performed, where the red dots signify single-mode states and the black dots signify multimode regimes. Below, four examples of numerically simulated regimes are depicted, along with their temporal evolution of normalized power (top) and optical spectrum (bottom): P1: CW. P2: Dense OFC. P3: HFC. P4: Irregular dynamics.
 } \label{LSA_ring_re}
\end{figure*}

We consider also the complex-conjugate of Eq.~(\ref{lsa2}):
\begin{align}
&iq{{{\delta a}_0}_n^*} +ik_n{{\delta a}_0}_n^*-i\omega{{\delta a}_0}_n^*+\lambda{{\ \delta a}_0}_n^*\nonumber\\ &=\sigma \Bigl[-{{\delta a}_0}_n^*+\left(1-i\alpha\right)\mu {{\ \delta a}_0}_n^* \nonumber\\&-2\left(1-i\alpha\right)\left( 2\left|a_0\right|^2{{\delta a}_0}_n^*+{a_0^*}^2{{{\delta a}_0}_{-n}} \right)\nonumber\\&+\left(\frac{1+i\alpha}{\mathrm{\Gamma}^2\left(1+\alpha^2\right)}\right) \left(-q^2{{\delta a}_0}_n^*+2iq\left(+ik_n\right){{\delta a}_0}_n^*-{k_n}^2{{\delta a}_0}_n^*\right)\Bigr]\label{lsa2cc}
\end{align}
Since $a_0$ can be assumed as a real number without loss of generality, as mentioned in Sec.~\ref{secsteady}, we rewrite Eqs.~(\ref{lsa2})-(\ref{lsa2cc}) as:
\begin{align}
{{\delta a}_0}_n\Bigl(W_n&+iY_n+i\omega+\lambda\Bigr)\nonumber\\&+{{{\delta a}_0}^{\phantom{-}\ast}_{-n}}\left(\sigma\left(1+i\alpha\right)\left|{a_0}\right|^2\right)=0\label{sec1}
\end{align}\begin{align}
{{\delta a}_0}_n\Bigl(\sigma\Bigl(1&-i\alpha\Bigr){\left|{a_0}\right|}^2\Bigr)\nonumber\\&+{{{\delta a}_0}^{\phantom{-}\ast}_{-n}}\left(W_{-n}-iY_{-n}-i\omega+\lambda\right)=0 \nonumber\\\label{sec2}
\end{align}

\begin{figure*} [t]
   \begin{center}
%   \begin{tabular}{c} %% tabular useful for creating an array of images 
   \includegraphics[width=0.8\textwidth]{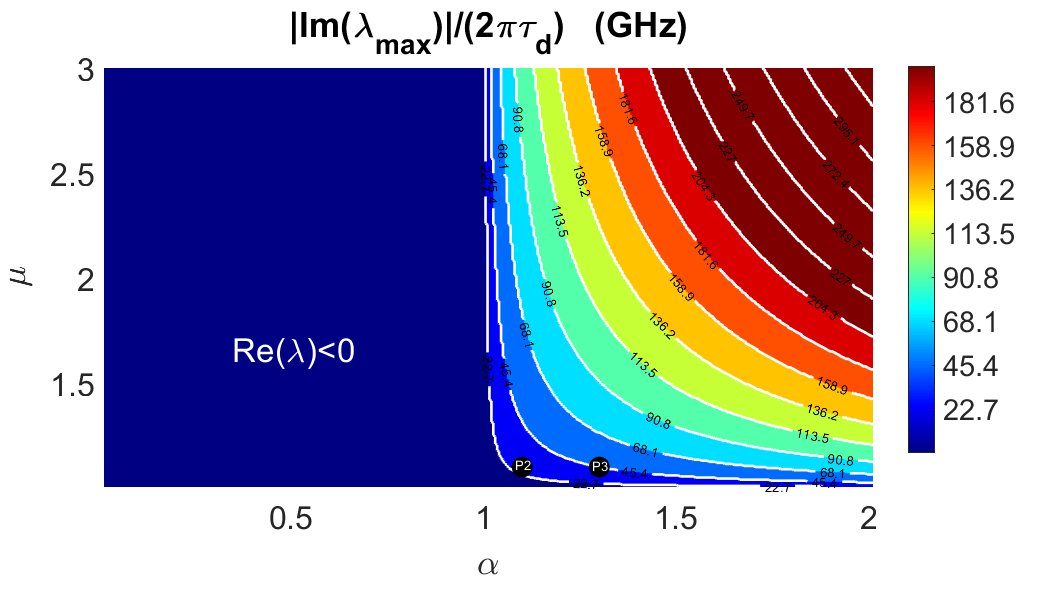}
%   \end{tabular}
   \end{center}
   \caption{Map showing the normalized imaginary part of $\lambda$ corresponding to the solution with the maximum real part, as a function of $\mu$ and $\alpha$ factor. The stability region of the CW solution ($\text{Re}(\lambda)<0$) is indicated by a dark blue color on the map. Points P2 and P3 in Fig.~\ref{LSA_ring_re}, corresponding to a dense OFC and an HFC, respectively, are represented as black dots on the map.
 } \label{LSA_ring_im}
\end{figure*}
\noindent where:
\begin{align}
W_n &=\sigma\Bigl[1-\mu+4\left|a_0\right|^2\nonumber\\ &+\frac{1}{\mathrm{\Gamma}^2\left(1+\alpha^2\right)}\left(q^2+2qk_n+{k_n}^2\right)\Bigr]\\
W_{-n}&=\sigma\Bigl[1-\mu+4\left|a_0\right|^2\nonumber\\ &+ \frac{1}{\mathrm{\Gamma}^2\left(1+\alpha^2\right)}\left(q^2-2qk_n+{k_n}^2\right)\Bigr]\\
%\end{align}\begin{align} uncomment to split in two pages
Y_n&=\sigma\Bigl[-\alpha\mu+2\alpha\left|a_0\right|^2\nonumber\\ &- \frac{\alpha}{\mathrm{\Gamma}^2\left(1+\alpha^2\right)}\left(q^2+2qk_n+{k_n}^2\right)\Bigr]\\
Y_{-n}&=\sigma\Bigl[-\alpha\mu+2\alpha\left|a_0\right|^2-\nonumber\\ &- \frac{\alpha}{\mathrm{\Gamma}^2\left(1+\alpha^2\right)}\left(q^2-2qk_n+{k_n}^2\right)\Bigr]
\end{align}
If $M$ is the characteristic matrix of the linear system composed by Eqs.~(\ref{sec1})-(\ref{sec2}) and we calculate $\mathrm{det}(M)=0$, we obtain:
\begin{align}
    &\left(W_n+iY_n+i\omega+\lambda\right)\left(W_{-n}-iY_{-n}-i\omega+\lambda\right)\nonumber\\&-\sigma^2\left(1+\alpha^2\right)\left|a_0\right|^4=0\label{secring}
\end{align}
Eq.~(\ref{secring}) is the secular equation.\\

\subsection{Numerical validation of the LSA}
We consider $\Gamma=0.09$, $L=4$ mm, $\sigma=4.5\times 10^{-4}$, and $\tau_d=0.1~$ps, and we solve Eq.~(\ref{secring}) for $\mu$ ranging from 1.005 to 3, and $\alpha$ ranging from 0.015 to 2, calculating the maximum value of $\text{Re}(\lambda)$ for each pair $(\alpha, \mu)$. Periodic boundary conditions have been applied in this study. Since $\mu_{\textrm{thr}}=1$, so that we can consider $\mu$ instead of the normalized pump parameter $p$.

The blue region on the map of Fig.~\ref{LSA_ring_re} corresponds to negative values of $\text{Re}(\lambda)$, indicating stable single-mode emission. To test the self-consistency of the model, we conducted numerical simulations, validating the predictions of the linear stability analysis. Some cases of performed simulations are represented by red and black dots on the map. The red dots, indicating single-mode states, are clustered in the region where $\text{Re}(\lambda)<0$, while the black dots (representing multimode simulated states) are found for $\text{Re}(\lambda)>0$, demonstrating consistency between linear stability analysis and numerical simulations. We particularly emphasize the consistency tested for $\alpha=1.1$, where we observe single-mode emission for $\mu=1.01$ (P1) and multimode emission for $\mu=1.1$ (P2). Notably, the dynamical regime observed in P2 corresponds to a fundamental OFC. Other instances of multimode dynamics, depicted as P3 and P4, illustrate a harmonic comb and irregular unlocked dynamics, respectively.

In Fig.~\ref{LSA_ring_re} we observe that initiating multimode dynamics requires an $\alpha$ value greater than 1. This implies that if the coupling between phase and field amplitude fails to reach a certain threshold (corresponding to $\alpha=1$ in this case), the system cannot develop multimode regimes and instead remains stable in single-mode emission regardless of the pump value. Additionally, we note that for $\alpha$ values between 1 and 1.2, distinct intervals of $\mu$ values can be discerned on the map, where the ring QCL displays single-mode behavior before transitioning to multimode as the $\mu$ increases. Conversely, for values of $\alpha$ greater than 1.2, the single-mode instability occurs very close to the laser threshold ($\mu=1$). These observations confirm the critical role of $\alpha$ in triggering multimode regimes, and consequently, in observing frequency combs, and are consistent with findings reported using the ESMBEs for the ring cavity in \cite{Columbo2018}. We also highlight that the transition from single-mode to multimode with increasing pump power has been experimentally observed in ring QCLs \cite{NaturePiccardo}, and the mentioned range of $\alpha$ values, where this behavior is more evident in the map of Fig.~\ref{LSA_ring_re}, are consistent with measurements reported in the literature for QCLs \cite{NaturePiccardo, Grillot16}.

We also investigate the role of the imaginary part of $\lambda$. Then, we consider the value of $\text{Im}(\lambda)$ corresponding to the maximum of $\text{Re}(\lambda)$ plotted in Fig.~\ref{LSA_ring_re}. After properly normalizing to have a quantity with the dimension of a frequency in GHz, we obtain the map in Fig.~\ref{LSA_ring_im}, where $\left|\text{Im}(\lambda_{\text{max}})\right|/(2\pi\tau d)$ is plotted as a function of $\mu$ and $\alpha$. The same case as in Fig.~\ref{LSA_ring_re} is considered. It should be noted that the values taken by $\left|\text{Im}(\lambda_{\text{max}})\right|/(2\pi\tau d)$ are integer multiples of the QCL cavity FSR. These values are organized into regions in the $\alpha$-$\mu$ plane, exhibiting boundaries shaped similarly to a hyperbola. We have verified that point P2 in Fig.~\ref{LSA_ring_re}, corresponding to a dense OFC with a spacing of 22.7~GHz, lies within the region where $\left|\text{Im}(\lambda_{\text{max}})\right|/(2\pi\tau d)$ achieves the specific value of 22.7~GHz. Furthermore, point P3 in Fig.~\ref{LSA_ring_im}, corresponding to a harmonic comb with a spacing of 45.4~GHz, exhibits a $\left|\text{Im}(\lambda_{\text{max}})\right|/(2\pi\tau d)$ value precisely equal to 45.4 GHz. This clearly demonstrates that the performed linear stability analysis not only identifies the stability region of the single-mode solution by examining the real part of $\lambda$, but also predicts the parameter space region where harmonic states can occur.
%\nocite{*}
\section{conclusion}\label{sec_concl}
We derived a single spatiotemporal equation to describe the multimodal dynamics of QCLs, following an order parameter approach. Our model serves as a powerful tool for investigating the physical properties of QCL combs, particularly facilitating comparative analyses between ring and FP configurations.

In this article, we emphasized the role of the integral term appearing in the equation, which reproduces the coupling between forward and backward fields in the FP cavity. The ability to gradually vary the impact of this term, by tuning a multiplicative coefficient K, allowed us to highlight how multiple localized structures, which manifest without coupling between the fields (ring), tend to disappear as K increases. This suggests the adverse impact of field coupling on the formation of harmonic combs.

This result is consistent with extensive numerical simulations, revealing a higher occurrence of harmonic combs in the ring configuration compared to FP. Furthermore, our simulation campaign successfully replicated typical temporal power and instantaneous frequency traces, facilitated by our model's ability to describe both amplitude and phase dynamics, representing a notable advancement from earlier models focusing on frequency modulation features \cite{Burghoff20}.

We note that the typical linear chirp characteristic of QCL OFCs is not observed for gain curve widths of a few hundred GHz, but it appears for larger values ($>1$ THz) of this parameter. This allows us to highlight how the observed linear chirp, present in experimental traces for mid-IR combs but not for THz combs, may be linked to the wider gain curve characteristic of mid-IR QCLs \cite{SilvestriReview,Faist_2016}.

Finally, we conducted a linear stability analysis for the single-mode solution in the ring case, leveraging the fact that these reduced models are particularly amenable to analytical treatment. We found that this tool holds strong predictive power not only regarding the destabilization of the CW solution but also for the formation of harmonic combs. This was observed through the analysis of the imaginary part of the exponent $\lambda$ within the perturbation expression.
 
\bibliography{apssamp}% Produces the bibliography via BibTeX.

\begin{thebibliography}{43}%
\makeatletter
\providecommand \@ifxundefined [1]{%
 \@ifx{#1\undefined}
}%
\providecommand \@ifnum [1]{%
 \ifnum #1\expandafter \@firstoftwo
 \else \expandafter \@secondoftwo
 \fi
}%
\providecommand \@ifx [1]{%
 \ifx #1\expandafter \@firstoftwo
 \else \expandafter \@secondoftwo
 \fi
}%
\providecommand \natexlab [1]{#1}%
\providecommand \enquote  [1]{``#1''}%
\providecommand \bibnamefont  [1]{#1}%
\providecommand \bibfnamefont [1]{#1}%
\providecommand \citenamefont [1]{#1}%
\providecommand \href@noop [0]{\@secondoftwo}%
\providecommand \href [0]{\begingroup \@sanitize@url \@href}%
\providecommand \@href[1]{\@@startlink{#1}\@@href}%
\providecommand \@@href[1]{\endgroup#1\@@endlink}%
\providecommand \@sanitize@url [0]{\catcode `\\12\catcode `\$12\catcode `\&12\catcode `\#12\catcode `\^12\catcode `\_12\catcode `\%12\relax}%
\providecommand \@@startlink[1]{}%
\providecommand \@@endlink[0]{}%
\providecommand \url  [0]{\begingroup\@sanitize@url \@url }%
\providecommand \@url [1]{\endgroup\@href {#1}{\urlprefix }}%
\providecommand \urlprefix  [0]{URL }%
\providecommand \Eprint [0]{\href }%
\providecommand \doibase [0]{https://doi.org/}%
\providecommand \selectlanguage [0]{\@gobble}%
\providecommand \bibinfo  [0]{\@secondoftwo}%
\providecommand \bibfield  [0]{\@secondoftwo}%
\providecommand \translation [1]{[#1]}%
\providecommand \BibitemOpen [0]{}%
\providecommand \bibitemStop [0]{}%
\providecommand \bibitemNoStop [0]{.\EOS\space}%
\providecommand \EOS [0]{\spacefactor3000\relax}%
\providecommand \BibitemShut  [1]{\csname bibitem#1\endcsname}%
\let\auto@bib@innerbib\@empty
%</preamble>
\bibitem [{\citenamefont {Faist}\ \emph {et~al.}(1994)\citenamefont {Faist}, \citenamefont {Capasso}, \citenamefont {Sivco}, \citenamefont {Sirtori}, \citenamefont {Hutchinson},\ and\ \citenamefont {Cho}}]{Faist_1994}%
  \BibitemOpen
  \bibfield  {author} {\bibinfo {author} {\bibfnamefont {J.}~\bibnamefont {Faist}}, \bibinfo {author} {\bibfnamefont {F.}~\bibnamefont {Capasso}}, \bibinfo {author} {\bibfnamefont {D.~L.}\ \bibnamefont {Sivco}}, \bibinfo {author} {\bibfnamefont {C.}~\bibnamefont {Sirtori}}, \bibinfo {author} {\bibfnamefont {A.~L.}\ \bibnamefont {Hutchinson}},\ and\ \bibinfo {author} {\bibfnamefont {A.~Y.}\ \bibnamefont {Cho}},\ }\bibfield  {title} {\bibinfo {title} {Quantum cascade laser},\ }\href@noop {} {\bibfield  {journal} {\bibinfo  {journal} {Science}\ }\textbf {\bibinfo {volume} {264}},\ \bibinfo {pages} {553} (\bibinfo {year} {1994})}\BibitemShut {NoStop}%
\bibitem [{\citenamefont {K{\"o}hler}\ \emph {et~al.}(2002)\citenamefont {K{\"o}hler}, \citenamefont {Tredicucci}, \citenamefont {Beltram}, \citenamefont {Beere}, \citenamefont {Linfield}, \citenamefont {Davies}, \citenamefont {Ritchie}, \citenamefont {Iotti},\ and\ \citenamefont {Rossi}}]{Kohler2002}%
  \BibitemOpen
  \bibfield  {author} {\bibinfo {author} {\bibfnamefont {R.}~\bibnamefont {K{\"o}hler}}, \bibinfo {author} {\bibfnamefont {A.}~\bibnamefont {Tredicucci}}, \bibinfo {author} {\bibfnamefont {F.}~\bibnamefont {Beltram}}, \bibinfo {author} {\bibfnamefont {H.~E.}\ \bibnamefont {Beere}}, \bibinfo {author} {\bibfnamefont {E.~H.}\ \bibnamefont {Linfield}}, \bibinfo {author} {\bibfnamefont {A.~G.}\ \bibnamefont {Davies}}, \bibinfo {author} {\bibfnamefont {D.~A.}\ \bibnamefont {Ritchie}}, \bibinfo {author} {\bibfnamefont {R.~C.}\ \bibnamefont {Iotti}},\ and\ \bibinfo {author} {\bibfnamefont {F.}~\bibnamefont {Rossi}},\ }\bibfield  {title} {\bibinfo {title} {Terahertz semiconductor-heterostructure laser},\ }\href@noop {} {\bibfield  {journal} {\bibinfo  {journal} {Nature}\ }\textbf {\bibinfo {volume} {417}},\ \bibinfo {pages} {156} (\bibinfo {year} {2002})}\BibitemShut {NoStop}%
\bibitem [{\citenamefont {Faist}\ \emph {et~al.}(2016)\citenamefont {Faist}, \citenamefont {Villares}, \citenamefont {Scalari}, \citenamefont {Rösch}, \citenamefont {Bonzon}, \citenamefont {Hugi},\ and\ \citenamefont {Beck}}]{Faist_2016}%
  \BibitemOpen
  \bibfield  {author} {\bibinfo {author} {\bibfnamefont {J.}~\bibnamefont {Faist}}, \bibinfo {author} {\bibfnamefont {G.}~\bibnamefont {Villares}}, \bibinfo {author} {\bibfnamefont {G.}~\bibnamefont {Scalari}}, \bibinfo {author} {\bibfnamefont {M.}~\bibnamefont {Rösch}}, \bibinfo {author} {\bibfnamefont {C.}~\bibnamefont {Bonzon}}, \bibinfo {author} {\bibfnamefont {A.}~\bibnamefont {Hugi}},\ and\ \bibinfo {author} {\bibfnamefont {M.}~\bibnamefont {Beck}},\ }\bibfield  {title} {\bibinfo {title} {Quantum cascade laser frequency combs},\ }\href {https://doi.org/doi:10.1515/nanoph-2016-0015} {\bibfield  {journal} {\bibinfo  {journal} {Nanophotonics}\ }\textbf {\bibinfo {volume} {5}},\ \bibinfo {pages} {272} (\bibinfo {year} {2016})}\BibitemShut {NoStop}%
\bibitem [{\citenamefont {Hugi}\ \emph {et~al.}(2012)\citenamefont {Hugi}, \citenamefont {Villares}, \citenamefont {Blaser}, \citenamefont {Liu},\ and\ \citenamefont {Faist}}]{Hugi2012}%
  \BibitemOpen
  \bibfield  {author} {\bibinfo {author} {\bibfnamefont {A.}~\bibnamefont {Hugi}}, \bibinfo {author} {\bibfnamefont {G.}~\bibnamefont {Villares}}, \bibinfo {author} {\bibfnamefont {S.}~\bibnamefont {Blaser}}, \bibinfo {author} {\bibfnamefont {H.~C.}\ \bibnamefont {Liu}},\ and\ \bibinfo {author} {\bibfnamefont {J.}~\bibnamefont {Faist}},\ }\bibfield  {title} {\bibinfo {title} {Mid-infrared frequency comb based on a quantum cascade laser},\ }\href {https://doi.org/10.1038/nature11620} {\bibfield  {journal} {\bibinfo  {journal} {Nature}\ }\textbf {\bibinfo {volume} {492}},\ \bibinfo {pages} {229} (\bibinfo {year} {2012})}\BibitemShut {NoStop}%
\bibitem [{\citenamefont {Burghoff}\ \emph {et~al.}(2014)\citenamefont {Burghoff}, \citenamefont {Kao}, \citenamefont {Han}, \citenamefont {Chan}, \citenamefont {Cai}, \citenamefont {Yang}, \citenamefont {Hayton}, \citenamefont {Gao}, \citenamefont {Reno},\ and\ \citenamefont {Hu}}]{Burghoff2014}%
  \BibitemOpen
  \bibfield  {author} {\bibinfo {author} {\bibfnamefont {D.}~\bibnamefont {Burghoff}}, \bibinfo {author} {\bibfnamefont {T.-Y.}\ \bibnamefont {Kao}}, \bibinfo {author} {\bibfnamefont {N.}~\bibnamefont {Han}}, \bibinfo {author} {\bibfnamefont {C.~W.~I.}\ \bibnamefont {Chan}}, \bibinfo {author} {\bibfnamefont {X.}~\bibnamefont {Cai}}, \bibinfo {author} {\bibfnamefont {Y.}~\bibnamefont {Yang}}, \bibinfo {author} {\bibfnamefont {D.~J.}\ \bibnamefont {Hayton}}, \bibinfo {author} {\bibfnamefont {J.-R.}\ \bibnamefont {Gao}}, \bibinfo {author} {\bibfnamefont {J.~L.}\ \bibnamefont {Reno}},\ and\ \bibinfo {author} {\bibfnamefont {Q.}~\bibnamefont {Hu}},\ }\bibfield  {title} {\bibinfo {title} {Terahertz laser frequency combs},\ }\href {https://doi.org/10.1038/nphoton.2014.85} {\bibfield  {journal} {\bibinfo  {journal} {Nature Photonics}\ }\textbf {\bibinfo {volume} {8}},\ \bibinfo {pages} {462} (\bibinfo {year} {2014})}\BibitemShut {NoStop}%
\bibitem [{\citenamefont {Piccardo}\ and\ \citenamefont {Capasso}(2022)}]{PiccardoReview}%
  \BibitemOpen
  \bibfield  {author} {\bibinfo {author} {\bibfnamefont {M.}~\bibnamefont {Piccardo}}\ and\ \bibinfo {author} {\bibfnamefont {F.}~\bibnamefont {Capasso}},\ }\bibfield  {title} {\bibinfo {title} {Laser frequency combs with fast gain recovery: Physics and applications},\ }\href@noop {} {\bibfield  {journal} {\bibinfo  {journal} {Laser \& Photonics Reviews}\ }\textbf {\bibinfo {volume} {16}},\ \bibinfo {pages} {2100403} (\bibinfo {year} {2022})}\BibitemShut {NoStop}%
\bibitem [{\citenamefont {Silvestri}\ \emph {et~al.}(2023)\citenamefont {Silvestri}, \citenamefont {Qi}, \citenamefont {Taimre}, \citenamefont {Bertling},\ and\ \citenamefont {Rakić}}]{SilvestriReview}%
  \BibitemOpen
  \bibfield  {author} {\bibinfo {author} {\bibfnamefont {C.}~\bibnamefont {Silvestri}}, \bibinfo {author} {\bibfnamefont {X.}~\bibnamefont {Qi}}, \bibinfo {author} {\bibfnamefont {T.}~\bibnamefont {Taimre}}, \bibinfo {author} {\bibfnamefont {K.}~\bibnamefont {Bertling}},\ and\ \bibinfo {author} {\bibfnamefont {A.~D.}\ \bibnamefont {Rakić}},\ }\bibfield  {title} {\bibinfo {title} {Frequency combs in quantum cascade lasers: An overview of modeling and experiments},\ }\href@noop {} {\bibfield  {journal} {\bibinfo  {journal} {APL Photonics}\ }\textbf {\bibinfo {volume} {8}},\ \bibinfo {pages} {020902} (\bibinfo {year} {2023})}\BibitemShut {NoStop}%
\bibitem [{\citenamefont {Burghoff}\ \emph {et~al.}(2015)\citenamefont {Burghoff}, \citenamefont {Yang}, \citenamefont {Hayton}, \citenamefont {Gao}, \citenamefont {Reno},\ and\ \citenamefont {Hu}}]{BurghoffSWIFTS}%
  \BibitemOpen
  \bibfield  {author} {\bibinfo {author} {\bibfnamefont {D.}~\bibnamefont {Burghoff}}, \bibinfo {author} {\bibfnamefont {Y.}~\bibnamefont {Yang}}, \bibinfo {author} {\bibfnamefont {D.~J.}\ \bibnamefont {Hayton}}, \bibinfo {author} {\bibfnamefont {J.-R.}\ \bibnamefont {Gao}}, \bibinfo {author} {\bibfnamefont {J.~L.}\ \bibnamefont {Reno}},\ and\ \bibinfo {author} {\bibfnamefont {Q.}~\bibnamefont {Hu}},\ }\bibfield  {title} {\bibinfo {title} {Evaluating the coherence and time-domain profile of quantum cascade laser frequency combs},\ }\href@noop {} {\bibfield  {journal} {\bibinfo  {journal} {Opt. Express}\ }\textbf {\bibinfo {volume} {23}},\ \bibinfo {pages} {1190} (\bibinfo {year} {2015})}\BibitemShut {NoStop}%
\bibitem [{\citenamefont {Singleton}\ \emph {et~al.}(2018)\citenamefont {Singleton}, \citenamefont {Jouy}, \citenamefont {Beck},\ and\ \citenamefont {Faist}}]{Optica_Faist_18}%
  \BibitemOpen
  \bibfield  {author} {\bibinfo {author} {\bibfnamefont {M.}~\bibnamefont {Singleton}}, \bibinfo {author} {\bibfnamefont {P.}~\bibnamefont {Jouy}}, \bibinfo {author} {\bibfnamefont {M.}~\bibnamefont {Beck}},\ and\ \bibinfo {author} {\bibfnamefont {J.}~\bibnamefont {Faist}},\ }\bibfield  {title} {\bibinfo {title} {Evidence of linear chirp in mid-infrared quantum cascade lasers},\ }\href {https://doi.org/10.1364/OPTICA.5.000948} {\bibfield  {journal} {\bibinfo  {journal} {Optica}\ }\textbf {\bibinfo {volume} {5}},\ \bibinfo {pages} {948} (\bibinfo {year} {2018})}\BibitemShut {NoStop}%
\bibitem [{\citenamefont {Meng}\ \emph {et~al.}(2020)\citenamefont {Meng}, \citenamefont {Singleton}, \citenamefont {Shahmohammadi}, \citenamefont {Kapsalidis}, \citenamefont {Wang}, \citenamefont {Beck},\ and\ \citenamefont {Faist}}]{Bomeng1}%
  \BibitemOpen
  \bibfield  {author} {\bibinfo {author} {\bibfnamefont {B.}~\bibnamefont {Meng}}, \bibinfo {author} {\bibfnamefont {M.}~\bibnamefont {Singleton}}, \bibinfo {author} {\bibfnamefont {M.}~\bibnamefont {Shahmohammadi}}, \bibinfo {author} {\bibfnamefont {F.}~\bibnamefont {Kapsalidis}}, \bibinfo {author} {\bibfnamefont {R.}~\bibnamefont {Wang}}, \bibinfo {author} {\bibfnamefont {M.}~\bibnamefont {Beck}},\ and\ \bibinfo {author} {\bibfnamefont {J.}~\bibnamefont {Faist}},\ }\bibfield  {title} {\bibinfo {title} {Mid-infrared frequency comb from a ring quantum cascade laser},\ }\href {https://doi.org/10.1364/OPTICA.377755} {\bibfield  {journal} {\bibinfo  {journal} {Optica}\ }\textbf {\bibinfo {volume} {7}},\ \bibinfo {pages} {162} (\bibinfo {year} {2020})}\BibitemShut {NoStop}%
\bibitem [{\citenamefont {Cappelli}\ \emph {et~al.}(2019)\citenamefont {Cappelli}, \citenamefont {Consolino}, \citenamefont {Campo}, \citenamefont {Galli}, \citenamefont {Mazzotti}, \citenamefont {Campa}, \citenamefont {Siciliani~de Cumis}, \citenamefont {Cancio~Pastor}, \citenamefont {Eramo}, \citenamefont {R{\"o}sch}, \citenamefont {Beck}, \citenamefont {Scalari}, \citenamefont {Faist}, \citenamefont {De~Natale},\ and\ \citenamefont {Bartalini}}]{Cappelli2019}%
  \BibitemOpen
  \bibfield  {author} {\bibinfo {author} {\bibfnamefont {F.}~\bibnamefont {Cappelli}}, \bibinfo {author} {\bibfnamefont {L.}~\bibnamefont {Consolino}}, \bibinfo {author} {\bibfnamefont {G.}~\bibnamefont {Campo}}, \bibinfo {author} {\bibfnamefont {I.}~\bibnamefont {Galli}}, \bibinfo {author} {\bibfnamefont {D.}~\bibnamefont {Mazzotti}}, \bibinfo {author} {\bibfnamefont {A.}~\bibnamefont {Campa}}, \bibinfo {author} {\bibfnamefont {M.}~\bibnamefont {Siciliani~de Cumis}}, \bibinfo {author} {\bibfnamefont {P.}~\bibnamefont {Cancio~Pastor}}, \bibinfo {author} {\bibfnamefont {R.}~\bibnamefont {Eramo}}, \bibinfo {author} {\bibfnamefont {M.}~\bibnamefont {R{\"o}sch}}, \bibinfo {author} {\bibfnamefont {M.}~\bibnamefont {Beck}}, \bibinfo {author} {\bibfnamefont {G.}~\bibnamefont {Scalari}}, \bibinfo {author} {\bibfnamefont {J.}~\bibnamefont {Faist}}, \bibinfo {author} {\bibfnamefont {P.}~\bibnamefont {De~Natale}},\ and\ \bibinfo {author} {\bibfnamefont {S.}~\bibnamefont {Bartalini}},\ }\bibfield  {title} {\bibinfo
  {title} {Retrieval of phase relation and emission profile of quantum cascade laser frequency combs},\ }\href@noop {} {\bibfield  {journal} {\bibinfo  {journal} {Nature Photonics}\ }\textbf {\bibinfo {volume} {13}},\ \bibinfo {pages} {562} (\bibinfo {year} {2019})}\BibitemShut {NoStop}%
\bibitem [{\citenamefont {Markmann}\ \emph {et~al.}(2023)\citenamefont {Markmann}, \citenamefont {Stark}, \citenamefont {Singleton}, \citenamefont {Beck}, \citenamefont {Faist},\ and\ \citenamefont {Scalari}}]{Faist_comb_profile_2023}%
  \BibitemOpen
  \bibfield  {author} {\bibinfo {author} {\bibfnamefont {S.}~\bibnamefont {Markmann}}, \bibinfo {author} {\bibfnamefont {D.}~\bibnamefont {Stark}}, \bibinfo {author} {\bibfnamefont {M.}~\bibnamefont {Singleton}}, \bibinfo {author} {\bibfnamefont {M.}~\bibnamefont {Beck}}, \bibinfo {author} {\bibfnamefont {J.}~\bibnamefont {Faist}},\ and\ \bibinfo {author} {\bibfnamefont {G.}~\bibnamefont {Scalari}},\ }\bibfield  {title} {\bibinfo {title} {Electro-optic sampling of a free-running terahertz quantum-cascade-laser frequency comb},\ }\href {https://doi.org/10.1103/PhysRevApplied.19.064063} {\bibfield  {journal} {\bibinfo  {journal} {Phys. Rev. Appl.}\ }\textbf {\bibinfo {volume} {19}},\ \bibinfo {pages} {064063} (\bibinfo {year} {2023})}\BibitemShut {NoStop}%
\bibitem [{\citenamefont {Piccardo}\ \emph {et~al.}(2018{\natexlab{a}})\citenamefont {Piccardo}, \citenamefont {Chevalier}, \citenamefont {Mansuripur}, \citenamefont {Kazakov}, \citenamefont {Wang}, \citenamefont {Rubin}, \citenamefont {Meadowcroft}, \citenamefont {Belyanin},\ and\ \citenamefont {Capasso}}]{PiccardoHFCOptex}%
  \BibitemOpen
  \bibfield  {author} {\bibinfo {author} {\bibfnamefont {M.}~\bibnamefont {Piccardo}}, \bibinfo {author} {\bibfnamefont {P.}~\bibnamefont {Chevalier}}, \bibinfo {author} {\bibfnamefont {T.~S.}\ \bibnamefont {Mansuripur}}, \bibinfo {author} {\bibfnamefont {D.}~\bibnamefont {Kazakov}}, \bibinfo {author} {\bibfnamefont {Y.}~\bibnamefont {Wang}}, \bibinfo {author} {\bibfnamefont {N.~A.}\ \bibnamefont {Rubin}}, \bibinfo {author} {\bibfnamefont {L.}~\bibnamefont {Meadowcroft}}, \bibinfo {author} {\bibfnamefont {A.}~\bibnamefont {Belyanin}},\ and\ \bibinfo {author} {\bibfnamefont {F.}~\bibnamefont {Capasso}},\ }\bibfield  {title} {\bibinfo {title} {The harmonic state of quantum cascade lasers: origin, control, and prospective applications},\ }\href@noop {} {\bibfield  {journal} {\bibinfo  {journal} {Opt. Express}\ }\textbf {\bibinfo {volume} {26}},\ \bibinfo {pages} {9464} (\bibinfo {year} {2018}{\natexlab{a}})}\BibitemShut {NoStop}%
\bibitem [{\citenamefont {Wang}\ \emph {et~al.}(2020)\citenamefont {Wang}, \citenamefont {Pistore}, \citenamefont {Riesch}, \citenamefont {Nong}, \citenamefont {Vigneron}, \citenamefont {Colombelli}, \citenamefont {Parillaud}, \citenamefont {Mangeney}, \citenamefont {Tignon}, \citenamefont {Jirauschek},\ and\ \citenamefont {Dhillon}}]{Dhillon1}%
  \BibitemOpen
  \bibfield  {author} {\bibinfo {author} {\bibfnamefont {F.}~\bibnamefont {Wang}}, \bibinfo {author} {\bibfnamefont {V.}~\bibnamefont {Pistore}}, \bibinfo {author} {\bibfnamefont {M.}~\bibnamefont {Riesch}}, \bibinfo {author} {\bibfnamefont {H.}~\bibnamefont {Nong}}, \bibinfo {author} {\bibfnamefont {P.-B.}\ \bibnamefont {Vigneron}}, \bibinfo {author} {\bibfnamefont {R.}~\bibnamefont {Colombelli}}, \bibinfo {author} {\bibfnamefont {O.}~\bibnamefont {Parillaud}}, \bibinfo {author} {\bibfnamefont {J.}~\bibnamefont {Mangeney}}, \bibinfo {author} {\bibfnamefont {J.}~\bibnamefont {Tignon}}, \bibinfo {author} {\bibfnamefont {C.}~\bibnamefont {Jirauschek}},\ and\ \bibinfo {author} {\bibfnamefont {S.~S.}\ \bibnamefont {Dhillon}},\ }\bibfield  {title} {\bibinfo {title} {Ultrafast response of harmonic modelocked {TH}z lasers},\ }\href@noop {} {\bibfield  {journal} {\bibinfo  {journal} {Light: Science {\&} Applications}\ }\textbf {\bibinfo {volume} {9}},\ \bibinfo {pages} {51} (\bibinfo {year} {2020})}\BibitemShut
  {NoStop}%
\bibitem [{\citenamefont {Forrer}\ \emph {et~al.}(2021)\citenamefont {Forrer}, \citenamefont {Wang}, \citenamefont {Beck}, \citenamefont {Belyanin}, \citenamefont {Faist},\ and\ \citenamefont {Scalari}}]{ForrerHFC}%
  \BibitemOpen
  \bibfield  {author} {\bibinfo {author} {\bibfnamefont {A.}~\bibnamefont {Forrer}}, \bibinfo {author} {\bibfnamefont {Y.}~\bibnamefont {Wang}}, \bibinfo {author} {\bibfnamefont {M.}~\bibnamefont {Beck}}, \bibinfo {author} {\bibfnamefont {A.}~\bibnamefont {Belyanin}}, \bibinfo {author} {\bibfnamefont {J.}~\bibnamefont {Faist}},\ and\ \bibinfo {author} {\bibfnamefont {G.}~\bibnamefont {Scalari}},\ }\bibfield  {title} {\bibinfo {title} {Self-starting harmonic comb emission in {TH}z quantum cascade lasers},\ }\href@noop {} {\bibfield  {journal} {\bibinfo  {journal} {Applied Physics Letters}\ }\textbf {\bibinfo {volume} {118}},\ \bibinfo {pages} {131112} (\bibinfo {year} {2021})}\BibitemShut {NoStop}%
\bibitem [{\citenamefont {Kazakov}\ \emph {et~al.}(2017)\citenamefont {Kazakov}, \citenamefont {Piccardo}, \citenamefont {Wang}, \citenamefont {Chevalier}, \citenamefont {Mansuripur}, \citenamefont {Xie}, \citenamefont {Zah}, \citenamefont {Lascola}, \citenamefont {Belyanin},\ and\ \citenamefont {Capasso}}]{Kazakov2017}%
  \BibitemOpen
  \bibfield  {author} {\bibinfo {author} {\bibfnamefont {D.}~\bibnamefont {Kazakov}}, \bibinfo {author} {\bibfnamefont {M.}~\bibnamefont {Piccardo}}, \bibinfo {author} {\bibfnamefont {Y.}~\bibnamefont {Wang}}, \bibinfo {author} {\bibfnamefont {P.}~\bibnamefont {Chevalier}}, \bibinfo {author} {\bibfnamefont {T.~S.}\ \bibnamefont {Mansuripur}}, \bibinfo {author} {\bibfnamefont {F.}~\bibnamefont {Xie}}, \bibinfo {author} {\bibfnamefont {C.-e.}\ \bibnamefont {Zah}}, \bibinfo {author} {\bibfnamefont {K.}~\bibnamefont {Lascola}}, \bibinfo {author} {\bibfnamefont {A.}~\bibnamefont {Belyanin}},\ and\ \bibinfo {author} {\bibfnamefont {F.}~\bibnamefont {Capasso}},\ }\bibfield  {title} {\bibinfo {title} {Self-starting harmonic frequency comb generation in a quantum cascade laser},\ }\href@noop {} {\bibfield  {journal} {\bibinfo  {journal} {Nature Photonics}\ }\textbf {\bibinfo {volume} {11}},\ \bibinfo {pages} {789} (\bibinfo {year} {2017})}\BibitemShut {NoStop}%
\bibitem [{\citenamefont {Kazakov}\ \emph {et~al.}(2021)\citenamefont {Kazakov}, \citenamefont {Opa\v{c}ak}, \citenamefont {Beiser}, \citenamefont {Belyanin}, \citenamefont {Schwarz}, \citenamefont {Piccardo},\ and\ \citenamefont {Capasso}}]{Kazakov2021}%
  \BibitemOpen
  \bibfield  {author} {\bibinfo {author} {\bibfnamefont {D.}~\bibnamefont {Kazakov}}, \bibinfo {author} {\bibfnamefont {N.}~\bibnamefont {Opa\v{c}ak}}, \bibinfo {author} {\bibfnamefont {M.}~\bibnamefont {Beiser}}, \bibinfo {author} {\bibfnamefont {A.}~\bibnamefont {Belyanin}}, \bibinfo {author} {\bibfnamefont {B.}~\bibnamefont {Schwarz}}, \bibinfo {author} {\bibfnamefont {M.}~\bibnamefont {Piccardo}},\ and\ \bibinfo {author} {\bibfnamefont {F.}~\bibnamefont {Capasso}},\ }\bibfield  {title} {\bibinfo {title} {Defect-engineered ring laser harmonic frequency combs},\ }\href {https://doi.org/10.1364/OPTICA.430896} {\bibfield  {journal} {\bibinfo  {journal} {Optica}\ }\textbf {\bibinfo {volume} {8}},\ \bibinfo {pages} {1277} (\bibinfo {year} {2021})}\BibitemShut {NoStop}%
\bibitem [{\citenamefont {Piccardo}\ \emph {et~al.}(2018{\natexlab{b}})\citenamefont {Piccardo}, \citenamefont {Kazakov}, \citenamefont {Rubin}, \citenamefont {Chevalier}, \citenamefont {Wang}, \citenamefont {Xie}, \citenamefont {Lascola}, \citenamefont {Belyanin},\ and\ \citenamefont {Capasso}}]{PiccardoSHB}%
  \BibitemOpen
  \bibfield  {author} {\bibinfo {author} {\bibfnamefont {M.}~\bibnamefont {Piccardo}}, \bibinfo {author} {\bibfnamefont {D.}~\bibnamefont {Kazakov}}, \bibinfo {author} {\bibfnamefont {N.~A.}\ \bibnamefont {Rubin}}, \bibinfo {author} {\bibfnamefont {P.}~\bibnamefont {Chevalier}}, \bibinfo {author} {\bibfnamefont {Y.}~\bibnamefont {Wang}}, \bibinfo {author} {\bibfnamefont {F.}~\bibnamefont {Xie}}, \bibinfo {author} {\bibfnamefont {K.}~\bibnamefont {Lascola}}, \bibinfo {author} {\bibfnamefont {A.}~\bibnamefont {Belyanin}},\ and\ \bibinfo {author} {\bibfnamefont {F.}~\bibnamefont {Capasso}},\ }\bibfield  {title} {\bibinfo {title} {Time-dependent population inversion gratings in laser frequency combs},\ }\href@noop {} {\bibfield  {journal} {\bibinfo  {journal} {Optica}\ }\textbf {\bibinfo {volume} {5}},\ \bibinfo {pages} {475} (\bibinfo {year} {2018}{\natexlab{b}})}\BibitemShut {NoStop}%
\bibitem [{\citenamefont {Silvestri}\ \emph {et~al.}(2020)\citenamefont {Silvestri}, \citenamefont {Columbo}, \citenamefont {Brambilla},\ and\ \citenamefont {Gioannini}}]{Silvestri20}%
  \BibitemOpen
  \bibfield  {author} {\bibinfo {author} {\bibfnamefont {C.}~\bibnamefont {Silvestri}}, \bibinfo {author} {\bibfnamefont {L.~L.}\ \bibnamefont {Columbo}}, \bibinfo {author} {\bibfnamefont {M.}~\bibnamefont {Brambilla}},\ and\ \bibinfo {author} {\bibfnamefont {M.}~\bibnamefont {Gioannini}},\ }\bibfield  {title} {\bibinfo {title} {Coherent multi-mode dynamics in a quantum cascade laser: amplitude- and frequency-modulated optical frequency combs},\ }\href {https://doi.org/10.1364/OE.396481} {\bibfield  {journal} {\bibinfo  {journal} {Opt. Express}\ }\textbf {\bibinfo {volume} {28}},\ \bibinfo {pages} {23846} (\bibinfo {year} {2020})}\BibitemShut {NoStop}%
\bibitem [{\citenamefont {Opa\v{c}ak}\ \emph {et~al.}(2021)\citenamefont {Opa\v{c}ak}, \citenamefont {Pilat}, \citenamefont {Kazakov}, \citenamefont {Cin}, \citenamefont {Ramer}, \citenamefont {Lendl}, \citenamefont {Capasso},\ and\ \citenamefont {Schwarz}}]{Opacakalpha}%
  \BibitemOpen
  \bibfield  {author} {\bibinfo {author} {\bibfnamefont {N.}~\bibnamefont {Opa\v{c}ak}}, \bibinfo {author} {\bibfnamefont {F.}~\bibnamefont {Pilat}}, \bibinfo {author} {\bibfnamefont {D.}~\bibnamefont {Kazakov}}, \bibinfo {author} {\bibfnamefont {S.~D.}\ \bibnamefont {Cin}}, \bibinfo {author} {\bibfnamefont {G.}~\bibnamefont {Ramer}}, \bibinfo {author} {\bibfnamefont {B.}~\bibnamefont {Lendl}}, \bibinfo {author} {\bibfnamefont {F.}~\bibnamefont {Capasso}},\ and\ \bibinfo {author} {\bibfnamefont {B.}~\bibnamefont {Schwarz}},\ }\bibfield  {title} {\bibinfo {title} {Spectrally resolved linewidth enhancement factor of a semiconductor frequency comb},\ }\href {https://doi.org/10.1364/OPTICA.428096} {\bibfield  {journal} {\bibinfo  {journal} {Optica}\ }\textbf {\bibinfo {volume} {8}},\ \bibinfo {pages} {1227} (\bibinfo {year} {2021})}\BibitemShut {NoStop}%
\bibitem [{\citenamefont {Vukovi\'c}\ \emph {et~al.}(2016)\citenamefont {Vukovi\'c}, \citenamefont {Radovanovi\'c}, \citenamefont {Milanovi\'c},\ and\ \citenamefont {Boiko}}]{Boiko1}%
  \BibitemOpen
  \bibfield  {author} {\bibinfo {author} {\bibfnamefont {N.}~\bibnamefont {Vukovi\'c}}, \bibinfo {author} {\bibfnamefont {J.}~\bibnamefont {Radovanovi\'c}}, \bibinfo {author} {\bibfnamefont {V.}~\bibnamefont {Milanovi\'c}},\ and\ \bibinfo {author} {\bibfnamefont {D.~L.}\ \bibnamefont {Boiko}},\ }\bibfield  {title} {\bibinfo {title} {Analytical expression for \uppercase{R}isken-\uppercase{N}ummedal-\uppercase{G}raham-\uppercase{H}aken instability threshold in quantum cascade lasers},\ }\href {https://doi.org/10.1364/OE.24.026911} {\bibfield  {journal} {\bibinfo  {journal} {Opt. Express}\ }\textbf {\bibinfo {volume} {24}},\ \bibinfo {pages} {26911} (\bibinfo {year} {2016})}\BibitemShut {NoStop}%
\bibitem [{\citenamefont {Vukovi\'c}\ \emph {et~al.}(2017)\citenamefont {Vukovi\'c}, \citenamefont {Radovanovi\'c}, \citenamefont {Milanovi\'c},\ and\ \citenamefont {Boiko}}]{Boiko2}%
  \BibitemOpen
  \bibfield  {author} {\bibinfo {author} {\bibfnamefont {N.~N.}\ \bibnamefont {Vukovi\'c}}, \bibinfo {author} {\bibfnamefont {J.}~\bibnamefont {Radovanovi\'c}}, \bibinfo {author} {\bibfnamefont {V.}~\bibnamefont {Milanovi\'c}},\ and\ \bibinfo {author} {\bibfnamefont {D.~L.}\ \bibnamefont {Boiko}},\ }\bibfield  {title} {\bibinfo {title} {Low-threshold \uppercase{RNGH} instabilities in quantum cascade lasers},\ }\href {https://doi.org/10.1109/JSTQE.2017.2699139} {\bibfield  {journal} {\bibinfo  {journal} {IEEE Journal of Selected Topics in Quantum Electronics}\ }\textbf {\bibinfo {volume} {23}},\ \bibinfo {pages} {1} (\bibinfo {year} {2017})}\BibitemShut {NoStop}%
\bibitem [{\citenamefont {Piccardo}\ \emph {et~al.}(2020)\citenamefont {Piccardo}, \citenamefont {Schwarz}, \citenamefont {Kazakov}, \citenamefont {Beiser}, \citenamefont {Opa{\v{c}}ak}, \citenamefont {Wang}, \citenamefont {Jha}, \citenamefont {Hillbrand}, \citenamefont {Tamagnone}, \citenamefont {Chen}, \citenamefont {Zhu}, \citenamefont {Columbo}, \citenamefont {Belyanin},\ and\ \citenamefont {Capasso}}]{NaturePiccardo}%
  \BibitemOpen
  \bibfield  {author} {\bibinfo {author} {\bibfnamefont {M.}~\bibnamefont {Piccardo}}, \bibinfo {author} {\bibfnamefont {B.}~\bibnamefont {Schwarz}}, \bibinfo {author} {\bibfnamefont {D.}~\bibnamefont {Kazakov}}, \bibinfo {author} {\bibfnamefont {M.}~\bibnamefont {Beiser}}, \bibinfo {author} {\bibfnamefont {N.}~\bibnamefont {Opa{\v{c}}ak}}, \bibinfo {author} {\bibfnamefont {Y.}~\bibnamefont {Wang}}, \bibinfo {author} {\bibfnamefont {S.}~\bibnamefont {Jha}}, \bibinfo {author} {\bibfnamefont {J.}~\bibnamefont {Hillbrand}}, \bibinfo {author} {\bibfnamefont {M.}~\bibnamefont {Tamagnone}}, \bibinfo {author} {\bibfnamefont {W.~T.}\ \bibnamefont {Chen}}, \bibinfo {author} {\bibfnamefont {A.~Y.}\ \bibnamefont {Zhu}}, \bibinfo {author} {\bibfnamefont {L.~L.}\ \bibnamefont {Columbo}}, \bibinfo {author} {\bibfnamefont {A.}~\bibnamefont {Belyanin}},\ and\ \bibinfo {author} {\bibfnamefont {F.}~\bibnamefont {Capasso}},\ }\bibfield  {title} {\bibinfo {title} {Frequency combs induced by phase turbulence},\ }\href
  {https://doi.org/10.1038/s41586-020-2386-6} {\bibfield  {journal} {\bibinfo  {journal} {Nature}\ }\textbf {\bibinfo {volume} {582}},\ \bibinfo {pages} {360} (\bibinfo {year} {2020})}\BibitemShut {NoStop}%
\bibitem [{\citenamefont {Columbo}\ \emph {et~al.}(2018)\citenamefont {Columbo}, \citenamefont {Barbieri}, \citenamefont {Sirtori},\ and\ \citenamefont {Brambilla}}]{Columbo2018}%
  \BibitemOpen
  \bibfield  {author} {\bibinfo {author} {\bibfnamefont {L.~L.}\ \bibnamefont {Columbo}}, \bibinfo {author} {\bibfnamefont {S.}~\bibnamefont {Barbieri}}, \bibinfo {author} {\bibfnamefont {C.}~\bibnamefont {Sirtori}},\ and\ \bibinfo {author} {\bibfnamefont {M.}~\bibnamefont {Brambilla}},\ }\bibfield  {title} {\bibinfo {title} {Dynamics of a broad-band quantum cascade laser: from chaos to coherent dynamics and mode-locking},\ }\href {http://opg.optica.org/oe/abstract.cfm?URI=oe-26-3-2829} {\bibfield  {journal} {\bibinfo  {journal} {Opt. Express}\ }\textbf {\bibinfo {volume} {26}},\ \bibinfo {pages} {2829} (\bibinfo {year} {2018})}\BibitemShut {NoStop}%
\bibitem [{\citenamefont {Friedli}\ \emph {et~al.}(2013)\citenamefont {Friedli}, \citenamefont {Sigg}, \citenamefont {Hinkov}, \citenamefont {Hugi}, \citenamefont {Riedi}, \citenamefont {Beck},\ and\ \citenamefont {Faist}}]{friedli}%
  \BibitemOpen
  \bibfield  {author} {\bibinfo {author} {\bibfnamefont {P.}~\bibnamefont {Friedli}}, \bibinfo {author} {\bibfnamefont {H.}~\bibnamefont {Sigg}}, \bibinfo {author} {\bibfnamefont {B.}~\bibnamefont {Hinkov}}, \bibinfo {author} {\bibfnamefont {A.}~\bibnamefont {Hugi}}, \bibinfo {author} {\bibfnamefont {S.}~\bibnamefont {Riedi}}, \bibinfo {author} {\bibfnamefont {M.}~\bibnamefont {Beck}},\ and\ \bibinfo {author} {\bibfnamefont {J.}~\bibnamefont {Faist}},\ }\bibfield  {title} {\bibinfo {title} {Four-wave mixing in a quantum cascade laser amplifier},\ }\href@noop {} {\bibfield  {journal} {\bibinfo  {journal} {Applied Physics Letters}\ }\textbf {\bibinfo {volume} {102}},\ \bibinfo {pages} {222104} (\bibinfo {year} {2013})}\BibitemShut {NoStop}%
\bibitem [{\citenamefont {Opa\ifmmode~\check{c}\else \v{c}\fi{}ak}\ and\ \citenamefont {Schwarz}(2019)}]{Opacak2019}%
  \BibitemOpen
  \bibfield  {author} {\bibinfo {author} {\bibfnamefont {N.}~\bibnamefont {Opa\ifmmode~\check{c}\else \v{c}\fi{}ak}}\ and\ \bibinfo {author} {\bibfnamefont {B.}~\bibnamefont {Schwarz}},\ }\bibfield  {title} {\bibinfo {title} {Theory of frequency-modulated combs in lasers with spatial hole burning, dispersion, and {K}err nonlinearity},\ }\href {https://doi.org/10.1103/PhysRevLett.123.243902} {\bibfield  {journal} {\bibinfo  {journal} {Phys. Rev. Lett.}\ }\textbf {\bibinfo {volume} {123}},\ \bibinfo {pages} {243902} (\bibinfo {year} {2019})}\BibitemShut {NoStop}%
\bibitem [{\citenamefont {Khurgin}\ \emph {et~al.}(2014)\citenamefont {Khurgin}, \citenamefont {Dikmelik}, \citenamefont {Hugi},\ and\ \citenamefont {Faist}}]{Khurgin2014}%
  \BibitemOpen
  \bibfield  {author} {\bibinfo {author} {\bibfnamefont {J.~B.}\ \bibnamefont {Khurgin}}, \bibinfo {author} {\bibfnamefont {Y.}~\bibnamefont {Dikmelik}}, \bibinfo {author} {\bibfnamefont {A.}~\bibnamefont {Hugi}},\ and\ \bibinfo {author} {\bibfnamefont {J.}~\bibnamefont {Faist}},\ }\bibfield  {title} {\bibinfo {title} {Coherent frequency combs produced by self frequency modulation in quantum cascade lasers},\ }\href@noop {} {\bibfield  {journal} {\bibinfo  {journal} {Applied Physics Letters}\ }\textbf {\bibinfo {volume} {104}},\ \bibinfo {pages} {081118} (\bibinfo {year} {2014})}\BibitemShut {NoStop}%
\bibitem [{\citenamefont {Tzenov}\ \emph {et~al.}(2016)\citenamefont {Tzenov}, \citenamefont {Burghoff}, \citenamefont {Hu},\ and\ \citenamefont {Jirauschek}}]{Tzenov2016}%
  \BibitemOpen
  \bibfield  {author} {\bibinfo {author} {\bibfnamefont {P.}~\bibnamefont {Tzenov}}, \bibinfo {author} {\bibfnamefont {D.}~\bibnamefont {Burghoff}}, \bibinfo {author} {\bibfnamefont {Q.}~\bibnamefont {Hu}},\ and\ \bibinfo {author} {\bibfnamefont {C.}~\bibnamefont {Jirauschek}},\ }\bibfield  {title} {\bibinfo {title} {Time domain modeling of terahertz quantum cascade lasers for frequency comb generation},\ }\href {http://opg.optica.org/oe/abstract.cfm?URI=oe-24-20-23232} {\bibfield  {journal} {\bibinfo  {journal} {Opt. Express}\ }\textbf {\bibinfo {volume} {24}},\ \bibinfo {pages} {23232} (\bibinfo {year} {2016})}\BibitemShut {NoStop}%
\bibitem [{\citenamefont {Jirauschek}\ and\ \citenamefont {Kubis}(2014)}]{Jiraus1}%
  \BibitemOpen
  \bibfield  {author} {\bibinfo {author} {\bibfnamefont {C.}~\bibnamefont {Jirauschek}}\ and\ \bibinfo {author} {\bibfnamefont {T.}~\bibnamefont {Kubis}},\ }\bibfield  {title} {\bibinfo {title} {{Modeling techniques for quantum cascade lasers}},\ }\href {https://doi.org/10.1063/1.4863665} {\bibfield  {journal} {\bibinfo  {journal} {Applied Physics Reviews}\ }\textbf {\bibinfo {volume} {1}},\ \bibinfo {pages} {011307} (\bibinfo {year} {2014})},\ \Eprint {https://arxiv.org/abs/https://pubs.aip.org/aip/apr/article-pdf/doi/10.1063/1.4863665/16702803/011307\_1\_online.pdf} {https://pubs.aip.org/aip/apr/article-pdf/doi/10.1063/1.4863665/16702803/011307\_1\_online.pdf} \BibitemShut {NoStop}%
\bibitem [{\citenamefont {Tzenov}\ \emph {et~al.}(2017)\citenamefont {Tzenov}, \citenamefont {Burghoff}, \citenamefont {Hu},\ and\ \citenamefont {Jirauschek}}]{Tzenov2}%
  \BibitemOpen
  \bibfield  {author} {\bibinfo {author} {\bibfnamefont {P.}~\bibnamefont {Tzenov}}, \bibinfo {author} {\bibfnamefont {D.}~\bibnamefont {Burghoff}}, \bibinfo {author} {\bibfnamefont {Q.}~\bibnamefont {Hu}},\ and\ \bibinfo {author} {\bibfnamefont {C.}~\bibnamefont {Jirauschek}},\ }\bibfield  {title} {\bibinfo {title} {Analysis of operating regimes of terahertz quantum cascade laser frequency combs},\ }\href {https://doi.org/10.1109/TTHZ.2017.2693822} {\bibfield  {journal} {\bibinfo  {journal} {IEEE Transactions on Terahertz Science and Technology}\ }\textbf {\bibinfo {volume} {7}},\ \bibinfo {pages} {351} (\bibinfo {year} {2017})}\BibitemShut {NoStop}%
\bibitem [{\citenamefont {Silvestri}\ \emph {et~al.}(2022)\citenamefont {Silvestri}, \citenamefont {Qi}, \citenamefont {Taimre},\ and\ \citenamefont {Raki\ifmmode~\acute{c}\else \'{c}\fi{}}}]{Silvestri22}%
  \BibitemOpen
  \bibfield  {author} {\bibinfo {author} {\bibfnamefont {C.}~\bibnamefont {Silvestri}}, \bibinfo {author} {\bibfnamefont {X.}~\bibnamefont {Qi}}, \bibinfo {author} {\bibfnamefont {T.}~\bibnamefont {Taimre}},\ and\ \bibinfo {author} {\bibfnamefont {A.~D.}\ \bibnamefont {Raki\ifmmode~\acute{c}\else \'{c}\fi{}}},\ }\bibfield  {title} {\bibinfo {title} {Multimode dynamics of terahertz quantum cascade lasers: Spontaneous and actively induced generation of dense and harmonic coherent regimes},\ }\href {https://doi.org/10.1103/PhysRevA.106.053526} {\bibfield  {journal} {\bibinfo  {journal} {Phys. Rev. A}\ }\textbf {\bibinfo {volume} {106}},\ \bibinfo {pages} {053526} (\bibinfo {year} {2022})}\BibitemShut {NoStop}%
\bibitem [{\citenamefont {Burghoff}(2020)}]{Burghoff20}%
  \BibitemOpen
  \bibfield  {author} {\bibinfo {author} {\bibfnamefont {D.}~\bibnamefont {Burghoff}},\ }\bibfield  {title} {\bibinfo {title} {Unraveling the origin of frequency modulated combs using active cavity mean-field theory},\ }\href {http://opg.optica.org/optica/abstract.cfm?URI=optica-7-12-1781} {\bibfield  {journal} {\bibinfo  {journal} {Optica}\ }\textbf {\bibinfo {volume} {7}},\ \bibinfo {pages} {1781} (\bibinfo {year} {2020})}\BibitemShut {NoStop}%
\bibitem [{\citenamefont {Humbard}\ and\ \citenamefont {Burghoff}(2022)}]{Humbard22}%
  \BibitemOpen
  \bibfield  {author} {\bibinfo {author} {\bibfnamefont {L.}~\bibnamefont {Humbard}}\ and\ \bibinfo {author} {\bibfnamefont {D.}~\bibnamefont {Burghoff}},\ }\bibfield  {title} {\bibinfo {title} {Analytical theory of frequency-modulated combs: generalized mean-field theory, complex cavities, and harmonic states},\ }\href@noop {} {\bibfield  {journal} {\bibinfo  {journal} {Opt. Express}\ }\textbf {\bibinfo {volume} {30}},\ \bibinfo {pages} {5376} (\bibinfo {year} {2022})}\BibitemShut {NoStop}%
\bibitem [{\citenamefont {Silvestri}(2022)}]{SilvestriThesis}%
  \BibitemOpen
  \bibfield  {author} {\bibinfo {author} {\bibfnamefont {C.}~\bibnamefont {Silvestri}},\ }\emph {\bibinfo {title} {Theory and modelization of Quantum Cascade Laser dynamics: comb formation, field structures and feedback-based imaging}},\ \href@noop {} {Ph.D. thesis},\ \bibinfo  {school} {Politecnico di Torino, Torino, Italy} (\bibinfo {year} {2022})\BibitemShut {NoStop}%
\bibitem [{\citenamefont {Opačak}\ \emph {et~al.}(2024)\citenamefont {Opačak}, \citenamefont {Kazakov}, \citenamefont {Columbo}, \citenamefont {Beiser}, \citenamefont {Letsou}, \citenamefont {Pilat}, \citenamefont {Brambilla}, \citenamefont {Prati}, \citenamefont {Piccardo}, \citenamefont {Capasso},\ and\ \citenamefont {Schwarz}}]{NB2024}%
  \BibitemOpen
  \bibfield  {author} {\bibinfo {author} {\bibfnamefont {N.}~\bibnamefont {Opačak}}, \bibinfo {author} {\bibfnamefont {D.}~\bibnamefont {Kazakov}}, \bibinfo {author} {\bibfnamefont {L.}~\bibnamefont {Columbo}}, \bibinfo {author} {\bibfnamefont {M.}~\bibnamefont {Beiser}}, \bibinfo {author} {\bibfnamefont {T.~P.}\ \bibnamefont {Letsou}}, \bibinfo {author} {\bibfnamefont {F.}~\bibnamefont {Pilat}}, \bibinfo {author} {\bibfnamefont {M.}~\bibnamefont {Brambilla}}, \bibinfo {author} {\bibfnamefont {F.}~\bibnamefont {Prati}}, \bibinfo {author} {\bibfnamefont {M.}~\bibnamefont {Piccardo}}, \bibinfo {author} {\bibfnamefont {F.}~\bibnamefont {Capasso}},\ and\ \bibinfo {author} {\bibfnamefont {B.}~\bibnamefont {Schwarz}},\ }\bibfield  {title} {\bibinfo {title} {Nozaki–{B}ekki solitons in semiconductor lasers},\ }\href@noop {} {\bibfield  {journal} {\bibinfo  {journal} {Nature}\ }\textbf {\bibinfo {volume} {625}},\ \bibinfo {pages} {685} (\bibinfo {year} {Jan 2024})}\BibitemShut {NoStop}%
\bibitem [{\citenamefont {Cole}\ \emph {et~al.}(2018)\citenamefont {Cole}, \citenamefont {Gatti}, \citenamefont {Papp}, \citenamefont {Prati},\ and\ \citenamefont {Lugiato}}]{Cole2018}%
  \BibitemOpen
  \bibfield  {author} {\bibinfo {author} {\bibfnamefont {D.~C.}\ \bibnamefont {Cole}}, \bibinfo {author} {\bibfnamefont {A.}~\bibnamefont {Gatti}}, \bibinfo {author} {\bibfnamefont {S.~B.}\ \bibnamefont {Papp}}, \bibinfo {author} {\bibfnamefont {F.}~\bibnamefont {Prati}},\ and\ \bibinfo {author} {\bibfnamefont {L.}~\bibnamefont {Lugiato}},\ }\bibfield  {title} {\bibinfo {title} {Theory of {Kerr} frequency combs in fabry-perot resonators},\ }\href {https://doi.org/10.1103/PhysRevA.98.013831} {\bibfield  {journal} {\bibinfo  {journal} {Phys. Rev. A}\ }\textbf {\bibinfo {volume} {98}},\ \bibinfo {pages} {013831} (\bibinfo {year} {2018})}\BibitemShut {NoStop}%
\bibitem [{\citenamefont {Campbell}\ \emph {et~al.}(2023)\citenamefont {Campbell}, \citenamefont {Hill}, \citenamefont {Del'Haye},\ and\ \citenamefont {Oppo}}]{Oppo2023}%
  \BibitemOpen
  \bibfield  {author} {\bibinfo {author} {\bibfnamefont {G.~N.}\ \bibnamefont {Campbell}}, \bibinfo {author} {\bibfnamefont {L.}~\bibnamefont {Hill}}, \bibinfo {author} {\bibfnamefont {P.}~\bibnamefont {Del'Haye}},\ and\ \bibinfo {author} {\bibfnamefont {G.-L.}\ \bibnamefont {Oppo}},\ }\bibfield  {title} {\bibinfo {title} {Dark solitons in {F}abry-{P}\'erot resonators with kerr media and normal dispersion},\ }\href {https://doi.org/10.1103/PhysRevA.108.033505} {\bibfield  {journal} {\bibinfo  {journal} {Phys. Rev. A}\ }\textbf {\bibinfo {volume} {108}},\ \bibinfo {pages} {033505} (\bibinfo {year} {2023})}\BibitemShut {NoStop}%
\bibitem [{\citenamefont {Prati}\ \emph {et~al.}(2021{\natexlab{a}})\citenamefont {Prati}, \citenamefont {Brambilla}, \citenamefont {Piccardo}, \citenamefont {Columbo}, \citenamefont {Silvestri}, \citenamefont {Gioannini}, \citenamefont {Gatti}, \citenamefont {Lugiato},\ and\ \citenamefont {Capasso}}]{Prati2021}%
  \BibitemOpen
  \bibfield  {author} {\bibinfo {author} {\bibfnamefont {F.}~\bibnamefont {Prati}}, \bibinfo {author} {\bibfnamefont {M.}~\bibnamefont {Brambilla}}, \bibinfo {author} {\bibfnamefont {M.}~\bibnamefont {Piccardo}}, \bibinfo {author} {\bibfnamefont {L.~L.}\ \bibnamefont {Columbo}}, \bibinfo {author} {\bibfnamefont {C.}~\bibnamefont {Silvestri}}, \bibinfo {author} {\bibfnamefont {M.}~\bibnamefont {Gioannini}}, \bibinfo {author} {\bibfnamefont {A.}~\bibnamefont {Gatti}}, \bibinfo {author} {\bibfnamefont {L.~A.}\ \bibnamefont {Lugiato}},\ and\ \bibinfo {author} {\bibfnamefont {F.}~\bibnamefont {Capasso}},\ }\bibfield  {title} {\bibinfo {title} {Soliton dynamics of ring quantum cascade lasers with injected signal},\ }\href {https://doi.org/doi:10.1515/nanoph-2020-0409} {\bibfield  {journal} {\bibinfo  {journal} {Nanophotonics}\ }\textbf {\bibinfo {volume} {10}},\ \bibinfo {pages} {195} (\bibinfo {year} {2021}{\natexlab{a}})}\BibitemShut {NoStop}%
\bibitem [{\citenamefont {Prati}\ \emph {et~al.}(2021{\natexlab{b}})\citenamefont {Prati}, \citenamefont {Lugiato}, \citenamefont {Gatti}, \citenamefont {Columbo}, \citenamefont {Silvestri}, \citenamefont {Gioannini}, \citenamefont {Brambilla}, \citenamefont {Piccardo},\ and\ \citenamefont {Capasso}}]{Pratichaos}%
  \BibitemOpen
  \bibfield  {author} {\bibinfo {author} {\bibfnamefont {F.}~\bibnamefont {Prati}}, \bibinfo {author} {\bibfnamefont {L.}~\bibnamefont {Lugiato}}, \bibinfo {author} {\bibfnamefont {A.}~\bibnamefont {Gatti}}, \bibinfo {author} {\bibfnamefont {L.}~\bibnamefont {Columbo}}, \bibinfo {author} {\bibfnamefont {C.}~\bibnamefont {Silvestri}}, \bibinfo {author} {\bibfnamefont {M.}~\bibnamefont {Gioannini}}, \bibinfo {author} {\bibfnamefont {M.}~\bibnamefont {Brambilla}}, \bibinfo {author} {\bibfnamefont {M.}~\bibnamefont {Piccardo}},\ and\ \bibinfo {author} {\bibfnamefont {F.}~\bibnamefont {Capasso}},\ }\bibfield  {title} {\bibinfo {title} {Global and localised temporal structures in driven ring quantum cascade lasers},\ }\href {https://doi.org/https://doi.org/10.1016/j.chaos.2021.111537} {\bibfield  {journal} {\bibinfo  {journal} {Chaos, Solitons \& Fractals}\ }\textbf {\bibinfo {volume} {153}},\ \bibinfo {pages} {111537} (\bibinfo {year} {2021}{\natexlab{b}})}\BibitemShut {NoStop}%
\bibitem [{\citenamefont {Columbo}\ \emph {et~al.}(2021)\citenamefont {Columbo}, \citenamefont {Piccardo}, \citenamefont {Prati}, \citenamefont {Lugiato}, \citenamefont {Brambilla}, \citenamefont {Gatti}, \citenamefont {Silvestri}, \citenamefont {Gioannini}, \citenamefont {Opa\ifmmode~\check{c}\else \v{c}\fi{}ak}, \citenamefont {Schwarz},\ and\ \citenamefont {Capasso}}]{Columbo2021}%
  \BibitemOpen
  \bibfield  {author} {\bibinfo {author} {\bibfnamefont {L.}~\bibnamefont {Columbo}}, \bibinfo {author} {\bibfnamefont {M.}~\bibnamefont {Piccardo}}, \bibinfo {author} {\bibfnamefont {F.}~\bibnamefont {Prati}}, \bibinfo {author} {\bibfnamefont {L.~A.}\ \bibnamefont {Lugiato}}, \bibinfo {author} {\bibfnamefont {M.}~\bibnamefont {Brambilla}}, \bibinfo {author} {\bibfnamefont {A.}~\bibnamefont {Gatti}}, \bibinfo {author} {\bibfnamefont {C.}~\bibnamefont {Silvestri}}, \bibinfo {author} {\bibfnamefont {M.}~\bibnamefont {Gioannini}}, \bibinfo {author} {\bibfnamefont {N.}~\bibnamefont {Opa\ifmmode~\check{c}\else \v{c}\fi{}ak}}, \bibinfo {author} {\bibfnamefont {B.}~\bibnamefont {Schwarz}},\ and\ \bibinfo {author} {\bibfnamefont {F.}~\bibnamefont {Capasso}},\ }\bibfield  {title} {\bibinfo {title} {Unifying frequency combs in active and passive cavities: Temporal solitons in externally driven ring lasers},\ }\href {https://doi.org/10.1103/PhysRevLett.126.173903} {\bibfield  {journal} {\bibinfo  {journal} {Phys. Rev.
  Lett.}\ }\textbf {\bibinfo {volume} {126}},\ \bibinfo {pages} {173903} (\bibinfo {year} {2021})}\BibitemShut {NoStop}%
\bibitem [{\citenamefont {Jumpertz}\ \emph {et~al.}(2016)\citenamefont {Jumpertz}, \citenamefont {Michel}, \citenamefont {Pawlus}, \citenamefont {Elsässer}, \citenamefont {Schires}, \citenamefont {Carras},\ and\ \citenamefont {Grillot}}]{Grillot16}%
  \BibitemOpen
  \bibfield  {author} {\bibinfo {author} {\bibfnamefont {L.}~\bibnamefont {Jumpertz}}, \bibinfo {author} {\bibfnamefont {F.}~\bibnamefont {Michel}}, \bibinfo {author} {\bibfnamefont {R.}~\bibnamefont {Pawlus}}, \bibinfo {author} {\bibfnamefont {W.}~\bibnamefont {Elsässer}}, \bibinfo {author} {\bibfnamefont {K.}~\bibnamefont {Schires}}, \bibinfo {author} {\bibfnamefont {M.}~\bibnamefont {Carras}},\ and\ \bibinfo {author} {\bibfnamefont {F.}~\bibnamefont {Grillot}},\ }\bibfield  {title} {\bibinfo {title} {Measurements of the linewidth enhancement factor of mid-infrared quantum cascade lasers by different optical feedback techniques},\ }\href@noop {} {\bibfield  {journal} {\bibinfo  {journal} {AIP Advances}\ }\textbf {\bibinfo {volume} {6}},\ \bibinfo {pages} {015212} (\bibinfo {year} {2016})}\BibitemShut {NoStop}%
\bibitem [{\citenamefont {Vitiello}\ and\ \citenamefont {Tredicucci}(2021)}]{Vitiellorev2}%
  \BibitemOpen
  \bibfield  {author} {\bibinfo {author} {\bibfnamefont {M.~S.}\ \bibnamefont {Vitiello}}\ and\ \bibinfo {author} {\bibfnamefont {A.}~\bibnamefont {Tredicucci}},\ }\bibfield  {title} {\bibinfo {title} {Physics and technology of terahertz quantum cascade lasers},\ }\href@noop {} {\bibfield  {journal} {\bibinfo  {journal} {Advances in Physics: X}\ }\textbf {\bibinfo {volume} {6}},\ \bibinfo {pages} {1893809} (\bibinfo {year} {2021})}\BibitemShut {NoStop}%
\bibitem [{\citenamefont {Li}\ \emph {et~al.}(2015)\citenamefont {Li}, \citenamefont {Laffaille}, \citenamefont {Gacemi}, \citenamefont {Apfel}, \citenamefont {Sirtori}, \citenamefont {Leonardon}, \citenamefont {Santarelli}, \citenamefont {R\"{o}sch}, \citenamefont {Scalari}, \citenamefont {Beck}, \citenamefont {Faist}, \citenamefont {H\"{a}nsel}, \citenamefont {Holzwarth},\ and\ \citenamefont {Barbieri}}]{Li15}%
  \BibitemOpen
  \bibfield  {author} {\bibinfo {author} {\bibfnamefont {H.}~\bibnamefont {Li}}, \bibinfo {author} {\bibfnamefont {P.}~\bibnamefont {Laffaille}}, \bibinfo {author} {\bibfnamefont {D.}~\bibnamefont {Gacemi}}, \bibinfo {author} {\bibfnamefont {M.}~\bibnamefont {Apfel}}, \bibinfo {author} {\bibfnamefont {C.}~\bibnamefont {Sirtori}}, \bibinfo {author} {\bibfnamefont {J.}~\bibnamefont {Leonardon}}, \bibinfo {author} {\bibfnamefont {G.}~\bibnamefont {Santarelli}}, \bibinfo {author} {\bibfnamefont {M.}~\bibnamefont {R\"{o}sch}}, \bibinfo {author} {\bibfnamefont {G.}~\bibnamefont {Scalari}}, \bibinfo {author} {\bibfnamefont {M.}~\bibnamefont {Beck}}, \bibinfo {author} {\bibfnamefont {J.}~\bibnamefont {Faist}}, \bibinfo {author} {\bibfnamefont {W.}~\bibnamefont {H\"{a}nsel}}, \bibinfo {author} {\bibfnamefont {R.}~\bibnamefont {Holzwarth}},\ and\ \bibinfo {author} {\bibfnamefont {S.}~\bibnamefont {Barbieri}},\ }\bibfield  {title} {\bibinfo {title} {Dynamics of ultra-broadband terahertz quantum cascade lasers for comb
  operation},\ }\href {https://doi.org/10.1364/OE.23.033270} {\bibfield  {journal} {\bibinfo  {journal} {Opt. Express}\ }\textbf {\bibinfo {volume} {23}},\ \bibinfo {pages} {33270} (\bibinfo {year} {2015})}\BibitemShut {NoStop}%
\end{thebibliography}%
\end{document}